\documentclass{article}

\usepackage{amsmath, amsthm, mathtools, amssymb, color, float, eqnarray, booktabs, makecell, soul, mdframed}
\usepackage[hyphens]{url}
\usepackage[square,numbers]{natbib}
\usepackage[justification=centering]{caption}
\usepackage{todonotes}
\usepackage{comment}
\usepackage{bm}
\usepackage{adjustbox}
\usepackage{subcaption} 

\pdfoutput=1

\usepackage[most]{tcolorbox}

\usepackage{graphicx, bm, amsmath, siunitx, longtable,tabularx, float, mathtools, booktabs, eqnarray, color, amsfonts, amssymb, appendix}
\usepackage{amsmath,amssymb,mathtools,eqnarray}  
\usepackage{empheq} 
\usepackage{upgreek}  
\usepackage{systeme}

\begin{document}


\title{The Capacity of Low Earth Orbit Computed using Source-sink Modeling}

\date{}

\author{Andrea D'Ambrosio\thanks{Postdoctoral Associate, Department of Aeronautics and Astronautics, Massachusetts Institute of Technology, Cambridge, MA 02139}, Miles Lifson\thanks{PhD student, Department of Aeronautics and Astronautics, Massachusetts Institute of Technology, Cambridge, MA 02139},
Richard Linares\thanks{Assistant Professor, Department of Aeronautics and Astronautics, Massachusetts Institute of Technology, Cambridge, MA 02139}
}

\maketitle
\begin{abstract}


The increasing number of Anthropogenic Space Objects (ASOs) in Low Earth Orbit (LEO) poses a threat to the safety and sustainability of the space environment. Multiple companies are planning to launch large constellations of hundreds or thousands of satellites in the near future, increasing congestion in LEO and the risk of collisions and debris generation. This paper employs a new multi-shell multi-species evolutionary source-sink model, called MOCAT-3, to estimate LEO orbital capacity. In particular, a new definition of orbital capacity based on the stable equilibrium points of the system is provided. Moreover, an optimization approach is used to compute the maximum orbital capacity of the low region of LEO (200-900 km of altitude), considering the equilibrium solutions and the failure rate of satellites as a constraint. Hence, an estimate for the maximum number of satellites that it is possible to fit in LEO, considering the stability of the space environment, is obtained. As a result, considering 7\% of failure rate, the maximum orbital capacity of LEO is estimated to be about 12.6 million satellites. Compatibility of future traffic launch, especially in terms of satellite constellations, is also analyzed and a strategy to accommodate for future traffic needs is proposed.

\end{abstract}
\section{Introduction}

In recent years, many companies have proposed large constellations, made up of hundreds or thousands of satellites, predominantly in Low Earth Orbit (LEO). Examples include the SpaceX Starlink constellation, with 4408 currently authorized satellites and a potential final number of about 42,000 satellites\footnote[1]{\url{https://licensing.fcc.gov/cgi-bin/ws.exe/prod/ib/forms/reports/swr031b.hts?q_set=V_SITE_ANTENNA_FREQ.file_numberC/File+Number/\%3D/SATAMD2021081800105&prepare=&column=V_SITE_ANTENNA_FREQ.file_numberC/File+Number}}, the Amazon Kuiper constellation, with about 3236 satellites authorized and a final number of about 7800 satellites\footnote[2]{\url{https://licensing.fcc.gov/cgi-bin/ws.exe/prod/ib/forms/reports/swr031b.hts?q_set=V_SITE_ANTENNA_FREQ.file_numberC/File+Number/\%3D/SATLOA2021110400145&prepare=&column=V_SITE_ANTENNA_FREQ.file_numberC/File+Number}}, and Astra Space constellation of about 13,000 satellites\footnote[3]{\url{https://licensing.fcc.gov/cgi-bin/ws.exe/prod/ib/forms/reports/swr031b.hts?q_set=V_SITE_ANTENNA_FREQ.file_numberC/File+Number/\%3D/SATLOA2021110400140&prepare=&column=V_SITE_ANTENNA_FREQ.file_numberC/File+Number}}. As a result of anticipated growth in the LEO Antropogenic Space Object (ASO) population, operators will face a denser and more complex operating environment characterized by both higher probabilities of collision and environmental consequences for any debris generating events.
Greater orbital density, from both more operational satellites and larger amounts of debris, poses challenges for space operators. In the short run, it imperils safety of flight and increases mission-ending risk. In the long run, a poorly stewarded LEO environment risks undermine its viability for current and future scientific, commercial, and national security uses.
For all these reasons, understanding the evolution of the space environment and the sensitivity of that evolution to different variables is essential in order to ensure the long-term sustainability of the space environment, and to inform work by operators to ensure kinetic space safety. 


This paper investigates the evolution of the ASO population in LEO by exploiting a new probabilistic source-sink model, which is part of the MIT Orbital Capacity Assessment Tool (MOCAT), with the objective of estimating LEO orbital capacity. In this work, orbital capacity refers to the number of satellites that is possible to fit in LEO. Understanding what is the maximum orbital capacity of LEO is essential to realize what is the current level of congestion of LEO, the available capacity for future launches, and to propose some strategies to accommodate for future needs. This analysis is carried out by means of the proposed source-sink model, which globally takes into account three object species, such as active satellites, derelict satellites, and debris. For this reason, we will refer to this model as MOCAT-3, which represents the three species version of the source-sink model developed under the MOCAT. The basics of the proposed source-sink model relies on the model introduced in the report written by the US government independent advisory body JASON and commissioned by the National Science Foundation\cite{jason}. That model has been improved by considering multiple altitude shells, different physical characteristics for the species involved, the NASA standard breakup model for catastrophic and non-catastrophic collisions \cite{krisko2011proper}, and a more precise evaluation of the natural orbital decay of the objects caused by the atmospheric drag.
Once introduced, the model is used to study the long-term evolution of ASOs. An estimation of LEO orbital capacity is provided according to a new proposed metric, associated with the equilibrium points of the dynamical system and the related stability over a long-term time period. The sensitivity of orbital capacity to launch rate variation is also investigated.

Generally, there are two different broad types of definitions of orbital capacity: intrinsic capacity of slotted spacecraft and risk-based capacity. Intrinsic capacity is related to the number and configuration of active ASOs that can be placed in a particular region of space in a self-compatible manner that avoids self-conjunctions indefinitely. It is a more space traffic management oriented definition that does not consider debris and other non-compliant ASOs. In contrast, risk-based capacity definitions typically incorporate representations of collisions and other stochastic phenomena to consider the evolution and distribution of ASOs over time \cite{Lifson2022}. There is not a single commonly-accepted metric to measure risk-based capacity, although various definitions have been proposed in literature. For example, the ECOB (Environmental Consequences of Orbital Breakups) index, which focuses on the evolution of the consequences of a fragmentation \cite{letizia2016assessment} and its likelihood \cite{letizia2017extending}. Other orbital capacity definitions are associated with the trend in number of fragments, such as the Number-Time product (NT) \cite{krag2017analysis}, or with indexes that measure the environmental impact of large bodies, such as the Criticality of the Spacecraft Index (CSI) \cite{rossi2015criticality}. To evaluate the CSI, many parameters are incorporated, including the ASO mass, lifetime, spatial density, and inclination. 

Many risk-based capacity metrics are derived from the evolution of ASOs over a certain time frame. In particular, two methodologies can be identified to estimate the temporal evolution of ASO populations. 
The first method exploits an approach where all the ASOs are propagated forward in time according to reasonably accurate physical models of spacecraft dynamics, often using semi-analytic propagation techniques for computational efficiency over long propagation times. This procedure typically takes into account many perturbations, such as atmospheric drag, solar radiation pressure, oblateness of the Earth, third-body perturbations, and space weather (including solar cycle effects). Collisions and explosions are often modeled as well. This approach allows computation of the exact information about individual ASOs. This methodology is often used in synergy with a Monte Carlo (MC) procedure, for which some uncertainties and stochastic input variables are added to study the stability and sensitivity of the space environment to different inputs. This allows for the computation of general statistics and probability density distributions for critical parameters. These models propagate single objects with greater accuracy because they model real physical dynamics for each ASO. However, they are computationally expensive and time consuming to run. Examples of models employing this approach are LEGEND \cite{liou2004legend}, DAMAGE \cite{lewis2001damage}, and DELTA \cite{walker2001analysis}.

The second methodology relies on the so-called source-sink models. These models are based on systems of coupled Ordinary Differential Equations (ODEs) describing the evolution of quantities of different species of ASOs, such as active payloads, rocket bodies, explosion and collision fragments, which can interact between each other \cite{somma2019adaptive}. Many discretizations can be introduced within those models, for example in terms of species, orbital altitude, and physical characteristics (mass, area, diameter). Moreover, objects can pass through multiple shells causing incoming and outgoing fluxes in the shells, for example because of the orbital decay due to the atmospheric drag. Other sources and sinks can be considered, such as new launches, post mission disposals, collisions, and explosions. This modeling approach surrenders the ability to study individual objects since objects are propagated as a species. In return, these models are computationally fast and can provide essential information about projected future distribution of ASOs in the space environment for long periods of time. Traditionally, these models can also be referred to as Particle-In-a-Box (PIB) models. Although these debris models represent a typical approach employed to study the long-term evolution of the LEO population, some simplifying assumptions are often carried out, thus leaving room for further improvements to make these models more realistic and reliable and calibrate them against either truth data or data from MC simulations. 
Many works in literature have already employed source-sink models to study the LEO population. As an example, Kessler and Cour-Palais developed a model which considered the major source and sink terms to study the evolution of the satellite population, predicting the significant growth of space debris due to collisions \cite{kessler1978collision}, with follow-on work by Kessler and Anz-Meador studying the stability of LEO using these modeling techniques based on empirical data and new break-up models \cite{Kessler2001}. Afterwards, the results of that work were revised and investigated again with more recent data and new models \cite{kessler2010kessler}, examining alternatives for controlling the future orbital debris environment. Furthermore, Talent \cite{talent1992analytic} proposed a simple model based on one first order ODE to describe the evolution trend of the objects in orbit. Somma et al. \cite{somma2017statistical} introduced a feedback controller within a statistical source sink-model to investigate adaptable debris control strategies. The model developed by Somma \cite{somma2019adaptive} and called MISSD (Model for Investigating control Strategies for Space Debris) has also been exploited by Trozzi et al. \cite{trozzi2021analysis} to study the evolution of the LEO region and analyze space environment capacity (including NT and the CSI).  

The main contributions of the current paper are the following:
\begin{itemize}
    \item A new source-sink model is introduced to study the evolution of ASOs.
    \item A new definition of the orbital capacity, directly related to the number of satellites that is possible to fit in LEO while maintaining the sustainability and stability of the space environment, is proposed.
    \item The maximum orbital capacity of LEO (within the range 200-900 km of altitude) is estimated, albeit without verification of the utilized model.
    \item An approach to assess compatibility of capacity with future traffic demands is demonstrated, along with a method to adjust shell populations to accommodate local demand above equilibrium by reducing populations elsewhere.
\end{itemize}

This manuscript is organized as follows. First, the MOCAT-3 model is introduced together with the main sources and sinks considered. Afterwards, the new definition of the orbital capacity is given and the approach pursued to obtain the maximum orbital capacity is explained. Hence, the overall framework is applied to estimate the maximum orbital capacity of LEO within the range 200-900 km of altitude. Several analysis, taking into account different failure rates and solar activity, are provided to study the sensitivity of the maximum orbital capacity to those parameters. The obtained solution is tested against the future traffic launches in terms of satellite constellations, and a strategy to accommodate for future needs is also proposed. Finally, concluding remarks and future research directions are given.

\section{MOCAT-3 Source-Sink Model}

\begin{figure}[!h]
    \centering
    \includegraphics[width=0.7\linewidth]{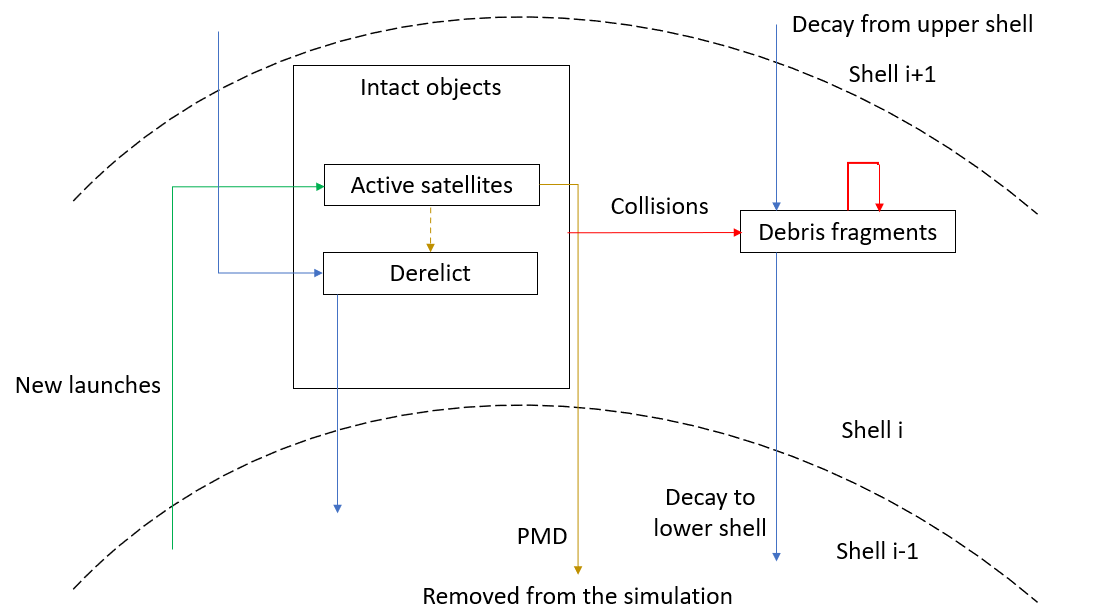}
    \caption{Schematics of the source-sink model.}
    \label{fig:old_model}
\end{figure}

The proposed source-sink model is a multi-bin multi-species model in which the LEO region is divided into many spherical orbital shells and three different species of objects are considered: active satellites ($S$), intact derelict satellites ($D$), which are intact but inactive satellites that fail to meet the post-mission disposal guidelines, and debris. 
The main assumptions employed to build the model are:
\begin{itemize}
    \item Near-circular orbits and non-rotating atmosphere are considered in the model.
    \item Solar radiation pressure is not accounted for in the model, since it has a minor effect in LEO. Moreover, Earth harmonics and third-body effects are not considered since only radial information about objects' positions are available using this model. Spherical symmetry of the earth gravitational potential is assumed and only atmospheric drag is considered as a perturbation. As a consequence, all the orbital elements of the ASOs population, but the semi-major axis, are not considered in the model. 
    \item New active satellites are assumed to be instantaneously injected into their final orbit altitude (no flows of objects from lower to upper shells are considered).
    \item Active satellites are not subject to the orbital decay effect caused by the atmospheric drag, since it is expected that they can perform station-keeping maneuvers to counteract the orbital decay and remain in their respective orbital shells.
    \item After the mean satellite lifetime ($\Delta t$), it is assumed that active satellites are directly removed from the simulation with a success rate of $P_M$, instead of performing maneuvers to move into elliptical or lower orbits (they do not change their orbital shell). This is more optimistic than assuming the compliance with the 25-year rule. However, it corresponds to what many companies have proposed for their large constellations.
    \item Explosions are not considered in the current model, but could be added in the future. Greater design reliability and technological developments are assumed in the future to minimize the frequency at which active satellites spontaneously explode.
    \item The minimum size of debris capable of disrupting a satellite is 10 cm, which is the minimum size that can be usually detectable by the Space Surveillance Network (SSN) and thus tracked for conjunction assessment and collision avoidance\footnote{\url{https://www.nasa.gov/mission_pages/station/news/orbital_debris.html}}. This assumption has been extensively employed in literature, such as in \cite{somma2019adaptive} and \cite{trozzi2021analysis}.
\end{itemize}
A schematics representing the qualitative interaction among the species is shown in Fig. \ref{fig:old_model}. 

Let $\bm{P}_h(t)=[S_h(t),D_h(t),N_h(t)]$ be the vector containing the whole population in the $h$-th altitude shell at time instant $t$. Please, note that hereafter the dependence from time $t$ and altitude (represented by the subscript $h$) will be removed for a better clarity in the notation; however, all the quantities are meant to be functions of time and altitude shell. The system of ODEs representing the evolution of the ASOs during time can be summarized by the following vector ODE:
\begin{equation}
    \label{eq:general_ODE}
    \dot{\bm{P}} = \dot{\bm{\Lambda}} + \dot{\bm{C}}_{PMD} + \dot{\bm{C}} + \dot{\bm{F}}
\end{equation}
where $\dot{\bm{\Lambda}}$ represents a term related to new launches, $\dot{\bm{C}}_{PMD}$ is related to Post-Mission Disposal, $\dot{\bm{C}}$ to collisions, and $\dot{\bm{F}}$ to other additional fluxes, such as atmospheric drag, which makes objects lower their altitude. As can be seen, the unit of Eq. \eqref{eq:general_ODE} is number of objects per year $\left[ \frac{\#}{year} \right]$. In the following part of this section, each term in Eq. \eqref{eq:general_ODE} is explained more in detail.

\paragraph{New Launches}

The term $\dot{\bm{\Lambda}}$ represents the launch rate expressed as the number of objects launched per year. This is the only source for the active satellites species. In fact, only active satellites (rather than debris or derelicts) are launched. Thus,
\begin{equation}
\label{eq:launches}
    \dot{\bm{\Lambda}}=[\lambda, 0 , 0]
\end{equation}
The $\lambda$ parameter can be used to account for future launch plans and to analyze the evolution of the LEO environment under the programmed or expected future launches.

\paragraph{Post-Mission Disposal}

As already stated, the Post-Mission Disposal (PMD) is implemented considering only the operative lifetime of satellites. This means that active satellites are removed from the simulation after $\Delta t$ years, with a probability of success equal to $P_{M}$. A percentage of the population of active satellites is assumed to fail to conduct PMD and becomes derelict. Therefore, 
\begin{equation}
\label{eq:PMD}
    \dot{\bm{C}}_{PMD} = \left[-\frac{S}{\Delta t}, +\frac{(1-P_M)S}{\Delta t},0 \right]
\end{equation}

The minus sign for the first component is related to the fact that active satellites are removed from the active satellites population $S$, whereas a percentage is added to the derelicts $D$ (plus sign). It is important to note that no active debris removal is considered in the model, meaning that atmospheric drag acts as the only sink to reduce debris levels.

\paragraph{Collisions}

The term $\dot{\bm{C}}$ related to the collisions among the species can be expanded as
\begin{equation}
    \label{eq:collisions}
    \dot{\bm{C}} = [\dot{C}_S,\dot{C}_D,\dot{C}_N \cdot n_f]
\end{equation}
where $n_f$ is the number of fragments generated by the collision. This quantity will be explained in greater detail later in this section.
In order to model collisions, the kinetic theory of gases is considered, an approach also implemented in previous works \cite{somma2017statistical,trozzi2021analysis}.
The components appearing in Eq. \eqref{eq:collisions} can be expressed as a function of the collision rate, as follows:
\begin{equation}
    \label{eq:collision_rate}
    \dot{C}_i = \sum_{j=1}^{N_{s}}\Gamma_{ij}\phi_{ij}Q_i Q_j + \dot{C}_{i,add}
\end{equation}
where $\dot{C}_{i,add}$ represents additional terms (introduced later in this section) corresponding to derelicts and debris generation, $N_s$ is the number of species considered (in this work $N_s=3$), $i,j=1,...,N_s$ are the subscripts indicating each species, $\Gamma_{ij}$ represent some coefficients related to collision avoidance probability and the derelict generation from a collision, and $\phi_{ij}$ is defined as:
\begin{equation}
    \label{eq:collision_intrinsic_prob}
    \phi_{ij} = \pi \frac{v_r(h) \sigma_{ij}}{V(h)}
\end{equation}
where $V(h)$ and $v_r(h)$ are the volume of the altitude shell and the average relative velocity in that shell. Even if average $v_r$ values vary with altitude, in this work it is assumed to be always equal to 10 km/s \cite{talent1992analytic}. Moreover, $\sigma_{ij}$ represents the square of the impact parameter and it is a function of the radius ($r$) of the colliding objects \cite{somma2019adaptive},
\begin{equation}
    \label{eq:impact_param}
    \sigma_{ij} = (r_i+r_j)^2
\end{equation}
The coefficients $\Gamma_{ij}$ can be thought as the elements of the following matrix $\Gamma$ (with $i$ and $j$ being the indexes of the rows and columns, respectively):
\begin{equation}
    \label{eq:Gamma_mat}
    \Gamma = \begin{bmatrix}
    -\alpha_a & -(\delta +\alpha) & -(\delta +\alpha) \\
    +\delta & -1 & -1 \\
    +\alpha & +\alpha & +1 
    \end{bmatrix}
\end{equation}
where $\delta$ is the ratio of the density of disabling to lethal debris (this term considers the possibility that collisions can generate new derelicts, other than debris), $\alpha$ and $\alpha_a$ are the fractions of collisions that an active satellite fails to avoid. In particular, $\alpha_a$ refers to the collisions among active satellites, whose occurrence is less frequent with respect to the collisions among the other species thanks to the collision avoidance maneuvers capability of active satellites ($\alpha_a<\alpha$).
The signs in the $\Gamma$ matrix actually indicate if the corresponding quantity is removed or added from a certain species. As can be seen, the last row associated to the debris has all positive components, since a collision always generates debris. The first component of the second row is positive since it is associated to the derelicts created from a collision.

Finally, the additional terms $\dot{C}_{i,add}$ are the components of the following vector $\dot{\bm{C}}_{add}$:
\begin{equation}
    \label{eq:C_add}
    \dot{\bm{C}}_{add} = [0, \, +\phi_{1,3}\delta SN, \, +\alpha_a \phi_{1,1}S^2+\phi_{1,2}\alpha SD+\phi_{2,2}D^2]
\end{equation}
These additional terms appear in the equations because we are considering the gain of derelicts due to the active satellites collisions, and the gain of debris corresponding to the active satellites events. 

The last parameter that needs to be computed is the number of fragments generated from a collision ($n_f$). This number is estimated using the NASA standard breakup models \cite{krisko2011proper}. Two different types of collisions are considered: catastrophic ($n_{f,c}$) and non-catastrophic/damaging ($n_{f,nc}$) collision. In this work, the collisions between intact objects ($S/D-S/D$) are assumed to be catastrophic, whereas the collisions with and between fragments ($N-S/D/N$) are considered non-catastrophic. The equations representing the two types of collision are:
\begin{gather}
    \label{eq:cat_col} 
    n_{f,c} = 0.1 \, L_C^{-1.71} (M_i+M_j)^{0.75}\\
    \label{eq:non_cat_col}
    n_{f,nc} = 0.1 \, L_C^{-1.71} (M_p\cdot v_{imp}^2)^{0.75}
\end{gather}
where $L_C$ is the characteristic length of the minimum size of generated debris (assumed to be 0.1 m), $M_{i/j}$ is the mass associated to the species $i/j$, $M_p$ is the mass of the projectile (i.e., the mass of the less massive object, $M_p=\min(M_i,M_j)$), and $v_{imp}$ is the impact velocity (supposed to be equal to 10 km/s).

\paragraph{Atmospheric Drag Effects}

The flux of objects decaying into lower altitude shells due to the atmospheric drag is the only flux considered in this paper. According to the assumptions stated before, since active satellites are not considered to be subject to the orbital decay because of the station-keeping maneuvers, we have
\begin{equation}
\label{eq:Drag}
    \dot{\bm{F}} = \left[0, \dot{F}_{d,D} , \dot{F}_{d,N}  \right]
\end{equation}
Indicating with $Q$ the number of objects belonging to a generic species, $\dot{F}_{d,Q}$ can be written as follows:
\begin{gather}
    \label{eq:SSM_Phi_d}
    \dot{F}_{d,Q} = - \frac{Q_{+}v_{+}}{d} + \frac{Q v}{d}
\end{gather}
with $d$ being the thickness of the shells (assumed to be spherical shells). The term $v$ is related to the change in semi-major axis, which is approximated with the radial distance $R$, and can be written as follows:
\begin{equation}
    \label{eq:v_semimajor_axis}
    v = -\rho B \sqrt{\mu R}
\end{equation}
where $\mu=398601$ km$^3$/s$^2$ is the Earth gravitational parameter, $B= c_D \frac{A}{m}$ is the ballistic coefficient with $c_D$, $A$ and $m$ being the drag coefficient, the area and the mass of the object, respectively. $\rho$ and $R$ are the atmospheric density and the distance computed from the Earth. The subscript ${+}$ refers to the quantities related to the upper shell; when it is not present, the quantities are meant to be computed in the current shell. The parameters that change according to the altitude shell are the number of objects $Q$, the atmospheric density $\rho$, and the distance $R$. Since different physical characteristics are taken into account for the species involved in the analysis, the ballistic coefficient varies according to the species considered. As can be seen, the unit of $\dot{F}_{d,Q}$ in Eq. \eqref{eq:SSM_Phi_d} is $\left[ \frac{\#}{time} \right]$, which is coherent with the unit of Eq. \eqref{eq:general_ODE}. 

Because atmospheric drag is the only natural sinking mechanism, as opposed to the artificial Post Mission Disposal and Active Debris Removal methods, it is important to employ an accurate model for the atmospheric density $\rho$. The density model that is currently employed in this work is a static exponential model \cite{vallado2013fundamentals}, obtained combining the U.S. Standard Atmosphere and the CIRA-72 model, 
\begin{equation}
    \label{eq:atm_density}
    \rho = \rho_0 \, exp{\left(-\frac{h-h_0}{H}\right)}
\end{equation}
where $\rho_0$ is the atmospheric density at reference altitude $h_0$, $H$ is the scale height, and $h$ is the altitude of the object. The values of these parameters can be found in Table 8-4 of Ref. \cite{vallado2013fundamentals}.

\paragraph{Final Model}

Combining all the previously presented terms, the time evolution of ASOs in each orbital shell for the 3-species proposed model is described by the following system of ODEs:
\begin{gather}
    \label{eq:SSM_S}
    \dot{S} = \lambda -S/\Delta t -\phi_{1,2}(\delta+\alpha)DS-\phi_{1,3}(\delta+\alpha)NS-\alpha_a\phi_{1,1} S^2 \\
    \label{eq:SSM_D}
    \dot{D} = \frac{(1-P_M)S}{\Delta t} + \phi_{1,2} \delta D S + \phi_{1,3} \delta N S  - \phi_{2,2} D^2 - \phi_{2,3} DN + \dot{F}_{d,D}\\
    \label{eq:SSM_N}
    \dot{N} = n_{f,13} \phi_{1,3} \alpha SN +n_{f,12} \phi_{1,2} \alpha SD +n_{f,22} \phi_{2,2} D^2+ \phi_{2,3} n_{f,23}DN + n_{f,11} \alpha_a \phi_{1,1} S^2 + \\ \nonumber +n_{f,33} \phi_{3,3} N^2 + \dot{F}_{d,N} 
\end{gather}
A better visualization of the interactions among the species and the parameters involved in the proposed model is provided in Table \ref{tab:SSM_interaction}.

\begin{table}[]
\centering
\begin{tabular}{|c|c|c|c|ccc|}
\hline
\textbf{}        & \textbf{\begin{tabular}[c]{@{}c@{}}$\mathbf{\dot{A}}$\\ New\\  Launches\end{tabular}} & \textbf{\begin{tabular}[c]{@{}c@{}}$\mathbf{\dot{C}_{pmd}}$\\ Post-\\ Mission\\ Disposal\end{tabular}} & \textbf{\begin{tabular}[c]{@{}c@{}}$\mathbf{\dot{F}}$\\ Drag\end{tabular}} & \multicolumn{3}{c|}{\textbf{\begin{tabular}[c]{@{}c@{}}$\mathbf{\dot{C}}$\\ Collision Source\end{tabular}}}                                                                                                                                                                                                                \\ \hline
\textbf{Species} &                                                                                       &                                                                                                        &                                                                            & \multicolumn{1}{c|}{\textbf{S}}                                                                                                                                       & \multicolumn{1}{c|}{\textbf{D}}                                                                                  & \textbf{N}                      \\ \hline
\textbf{S}       & $\lambda$                                                                             & $\frac{S}{\Delta t}$                                                                                   & 0                                                                          & \multicolumn{1}{c|}{$-\alpha_a \phi_{1,1}S^2$}                                                                                                                        & \multicolumn{1}{c|}{$-\phi_{1,2}(\delta + \alpha) SD$}                                                           & $-\phi_{1,3}(\delta+\alpha)S N$ \\
\hline
\textbf{D}       & 0                                                                                     & $\frac{(1-P_m)S}{\Delta t}$                                                                            & $\dot{F_{d,D}}$                                                            & \multicolumn{1}{c|}{\begin{tabular}[c]{@{}c@{}}$\phi_{1,2}\delta D S $\\ $+\phi_{1,3}\delta N S $\end{tabular}}                                                       & \multicolumn{1}{c|}{$-\phi_{2,2}D^2$}                                                                            & $ -\phi_{2,3} DN $              \\ \hline
\textbf{N}       & 0                                                                                     & 0                                                                                                      & $\dot{F_{d,N}}$                                                            & \multicolumn{1}{c|}{\begin{tabular}[c]{@{}c@{}}$n_{f,13} \phi_{1,3}\alpha S N$\\ $+n_{f,12} \phi_{1,2} \alpha SD$\\ $+n_{f,11} \alpha_a \phi_{1,1} S^2$\end{tabular}} & \multicolumn{1}{c|}{\begin{tabular}[c]{@{}c@{}}$n_{f,22} \phi_{2,2} D^2$\\ $+\phi_{2,3}n_{f,23} DN$\end{tabular}} & $n_{f,33}\phi_{3,3}N^2$         \\ \hline
\end{tabular}
\caption{Interactions among the species of the MOCAT-3.}
\label{tab:SSM_interaction}
\end{table}


\section{Orbital Capacity}\label{sec: Orb_cap}

There is significant need for methods to estimate the capacity of a particular region of space. As already stated in the introduction, there is not a commonly-accepted definition for orbital capacity within the scientific community. Many definitions and indexes have been proposed, such as the ECOB index \cite{letizia2016assessment,letizia2017extending}, the Number-Time product \cite{krag2017analysis}, and the CSI \cite{rossi2015criticality}. However, these indexes are not directly related to the number of satellites that it is actually possible to fit in LEO while preserving sustainability.

A new definition of orbital capacity is proposed in this paper, based on dynamical system analysis, to estimate the permissible number of satellites directly subject to a long-term stability constraint. In order to do that, bounded solutions that can ensure the stability of the space environment should be considered. Specifically, one possibility to do that is to compute the equilibrium points of the dynamical system represented by Eqs. \eqref{eq:general_ODE} or \eqref{eq:SSM_S}-\eqref{eq:SSM_N}. One advantage of performing dynamical analysis and relying on equilibrium points for the capacity is that it is independent from any initial conditions related to the initial ASOs population and the period of time of the analysis, which on the other hand would provide results just for specific cases. The proposed orbital capacity metrics ($\chi$), referred to as $\chi$-capacity throughout the paper for convenience, is defined as follows:
\begin{equation}
    \label{eq:orbital_capacity}
    \chi = \frac{S_{NC}-S_{eq}}{S_{NC}}
\end{equation}
where $S_{NC}=\lambda \Delta t$ represents the ideal number of satellites when all active satellites perfectly avoid collisions and there are no derelicts or debris, and $S_{eq}$ is the number of satellites corresponding to an equilibrium point of Eq. \eqref{eq:general_ODE} considering the full model with debris and collisions. The proposed capacity definition relates the number of active satellites at equilibrium between the real world (with collisions and debris) and the ideal world (without collisions and debris). As a further remark, the attentive reader can notice that the $\chi$-capacity can never be zero, since our current technology does not allow to always avoid collisions and prevent the generation of new debris.

As already stated, the $\chi$-capacity relies on the computation of the equilibrium points of the dynamical system. This can be carried out for example by using symbolic toolboxes. However, computing the equilibrium points of that system when multiple shells are considered can be quite cumbersome, since the number of equations and unknowns in the system increases. At this point, it is important to notice that, according to our hypothesis, no flows going from lower to upper shells are considered and the population in each shell ($\bm{P}_h(t)$) depends just on the quantities of the shells above. In particular, the only variables that are needed as inputs from the above shells are $D_{h+1}(t)$ and $N_{h+1}(t)$, where the subscript $h$ represents the $h$-th shell. This means that it is possible to compute the equilibrium points of each shell consecutively starting from the highest shell and going downward. Indeed, if an equilibrium is reached in one shell, it means that the amount of derelicts and debris decaying to the lower shell is constant. 

When solving for the equilibrium points, the populations $S_h$, $D_h$, and $N_h$ are considered unknown, whereas the launch rate is supposed to be known and fixed. Launch rate is assumed to be known since we are interested in obtaining the equilibrium solution given a certain defined launch rate. However, if we are interested in optimizing the launch rate to minimize/maximize a cost function (for example to maximize orbital capacity, as it is explained later on in this section), another outside optimization loop is required to compute the optimal launch rate. In this case, the optimizer varies the launch rate, which is passed as a fixed and known variable to the function that computes the equilibrium points and the associated orbital capacity.

In order to compute the equilibrium points of Eqs. \eqref{eq:SSM_S}-\eqref{eq:SSM_N}, the derivatives [$\dot{S}$, $\dot{D}$, $\dot{N}$] must be set equal to zero. This leads to obtain three second order polynomials equations, yielding an eighth-ordered system. The eight corresponding equilibrium points are found symbolically. However, the only acceptable and meaningful solutions are the non-negative solutions (excluding the trivial null solution). 

\subsection{Maximum Orbital Capacity}

In order to obtain the maximum orbital capacity that leads to a sustainable and safe use of LEO, considering the definition given in Eq. \eqref{eq:orbital_capacity}, an optimization approach can be carried out to compute the launch rate that maximizes the number of active satellites that we can fit in LEO. Therefore, the following single objective cost function to minimize, including the term related to the constraints, can be written as:
\begin{equation}
    \label{eq:cost_function}
    J = \frac{\beta_1}{S_{tot,eq}}+\beta_2 \sum_{i=1}^{N_{vin}} (S_{i,NC}-S_{i,eq})
\end{equation}
where $S_{tot,eq}$ is the total number of active satellites computed at the equilibrium point considering the entire system of ODEs, $\beta_1>0$ and $\beta_2\geq 0$ are two weights that balance the real objective of the cost function and the constraints to satisfy, represented by the second term of Eq. \eqref{eq:cost_function}, and $N_{vin}$ is the number of shells where the following constraint, evaluated for each shell, is not satisfied:
\begin{equation}
    \label{eq:chi_bar}
    \chi \leq \bar{\chi}
\end{equation}
with $\bar{\chi}$ being a tolerance percentage set by the user, here referred to as the failure rate. This failure rate basically takes into account the allowed percentage of satellites failure due to collisions and the presence of space debris. 
$\beta_2$ can be written as:
\begin{equation}
    \label{eq:constraints}
    \beta_2 = \begin{cases}
    0 \qquad \textrm{if} \; \chi \leq \bar{\chi} \; (\textrm{for each shell}) \\
    \beta_{22} \qquad \textrm{otherwise}
    \end{cases}
\end{equation}
where $\beta_{22}$ is a positive constant.

One can observe that, in order to compute the maximum capacity, the equilibrium point ($S_{eq}$) chosen among the eight feasible solutions is the one with the highest number of active satellites.

\section{Numerical Results}\label{sec: Results}

In this section, numerical results are discussed. Before these results are presented, we caution that MOCAT-3 has not yet been verified versus historical or higher fidelity modeling data. Many parameters have been set to either match the literature or simply be reasonable, but may not necessarily correspond to trends in the actual space environment. Results are presented to demonstrate the modeling and analytical approaches proposed, but should not be interpreted as providing conclusive statements regarding the behavior of the actual LEO orbital environment. Verification is intended in future work.

The region of LEO that is analyzed in this work ranges from $h_{min}=200$ km to $h_{max} = 900$ km of altitude divided into $N_{shells}$ orbital shells. The parameters employed for the simulations are reported in Table \ref{tab:model_param}. In order to compute the equilibrium points just for the highest shell, we assume that the incoming number of derelicts and debris coming from above is the same as the number of derelicts and debris of the current shell, i.e., $D_{h+1}=D_h$ and $N_{h+1}=N_h$. The physical characteristics of the ASOs are shown in Table \ref{tab:phys_carac}, where the same average values of mass, diameter and area ($m$, $b$, and $A$ respectively) are considered for the objects belonging to the same species. These values are taken from the average values reported in \cite{somma2019adaptive}. It is worth to notice that derelicts present the same values as active satellites, since the derelict family is composed of active satellites which fail in the PMD and become inactive. The probability of success of the PMD is chosen to be $P_M=95\%$ with an operational lifetime of active satellites of $\Delta t = 5$ years. As already stated, the relative impact velocity considered for the collision is a constant value set equal to $v_r = 10$ km/s, which is a common value employed in literature \cite{somma2019adaptive}. The parameters $\alpha$ and $\delta$ mirror those used in the JASON report for comparability purposes \cite{jason}. The minimum size of the fragments generated from a collision is $L_C = 0.1$ m, which also corresponds to the generally-accepted minimum size of detectable debris. 

\begin{table}[htb!]
\small
    \centering
    \begin{tabular}{cccccccccccc}
    \hline
    $h_{min}$ & $h_{max}$ & $N_{shells}$ & $d$ & $\Delta t$ & $v_r$ & $\alpha$ & $\alpha_a$ & $\delta$ & $P_M$ & $c_D$ & $L_C$  \\
    \hline
    200 km & 900 km & 20 & 35 km & 5 years & 10 km/s & 0.2 & 0.01 & 10 & 95\% & 2.2 & 0.1 m \\
    \hline    
    \end{tabular}
    \caption{Parameters employed for the simulations.}
    \label{tab:model_param}
\end{table}

\begin{table}[htb!]
\small
    \centering
    \begin{tabular}{cccc}
    \hline
    ASO & m [kg] & $b$ [m] & $A$ [m$^2$]\\
    \hline
    S & 223 & 1.49 & 1.741 \\
    D & 223 & 1.49 & 1.741 \\
    N & 0.64 & 0.18 & 0.02 \\
    \hline    
    \end{tabular}
    \caption{Average physical characteristics of the ASOs.}
    \label{tab:phys_carac}
\end{table}

The optimization is carried out by means of the Particle Swarm Optimization (PSO) metaheuristic algorithm. The cost function to minimize is expressed by Eq. \eqref{eq:cost_function}, where $\beta_1 = 10^6$ and $\beta_{22}=10^3$. 
A different launch rate for each shell is considered, which means that $N_{shells}$ optimization variables are actually taken into account. By doing this, the optimizer has more freedom to choose a possibly (very) different value of launch rate for each shell, encouraging a higher number of launches in lower shells, as can be expected. Instead of performing the optimization on all the shells at the same time, it was noticed that optimizing each shell separately from the others actually provides better solutions. Hence, we start the optimization from the highest shell and then we continue downwards, using the outputs of each shell as inputs for the shells below. Once the solution is achieved, the chosen optimal equilibrium point of each shell is collected into a matrix, so that to test the overall solution all at once within the multiple shells model. This is useful to check any possible differences that can arise by considering the model shell by shell (as done with the optimization) or all the shells together (for example during the long-term propagation). Hence, the stability of the obtained solution, given the computed equilibrium point and the optimal launch rate for each shell, can be studied by writing the Jacobian matrix of the overall system and computing its eigenvalues.

\begin{figure}[!h]
    \centering
    \includegraphics[width=0.6\linewidth]{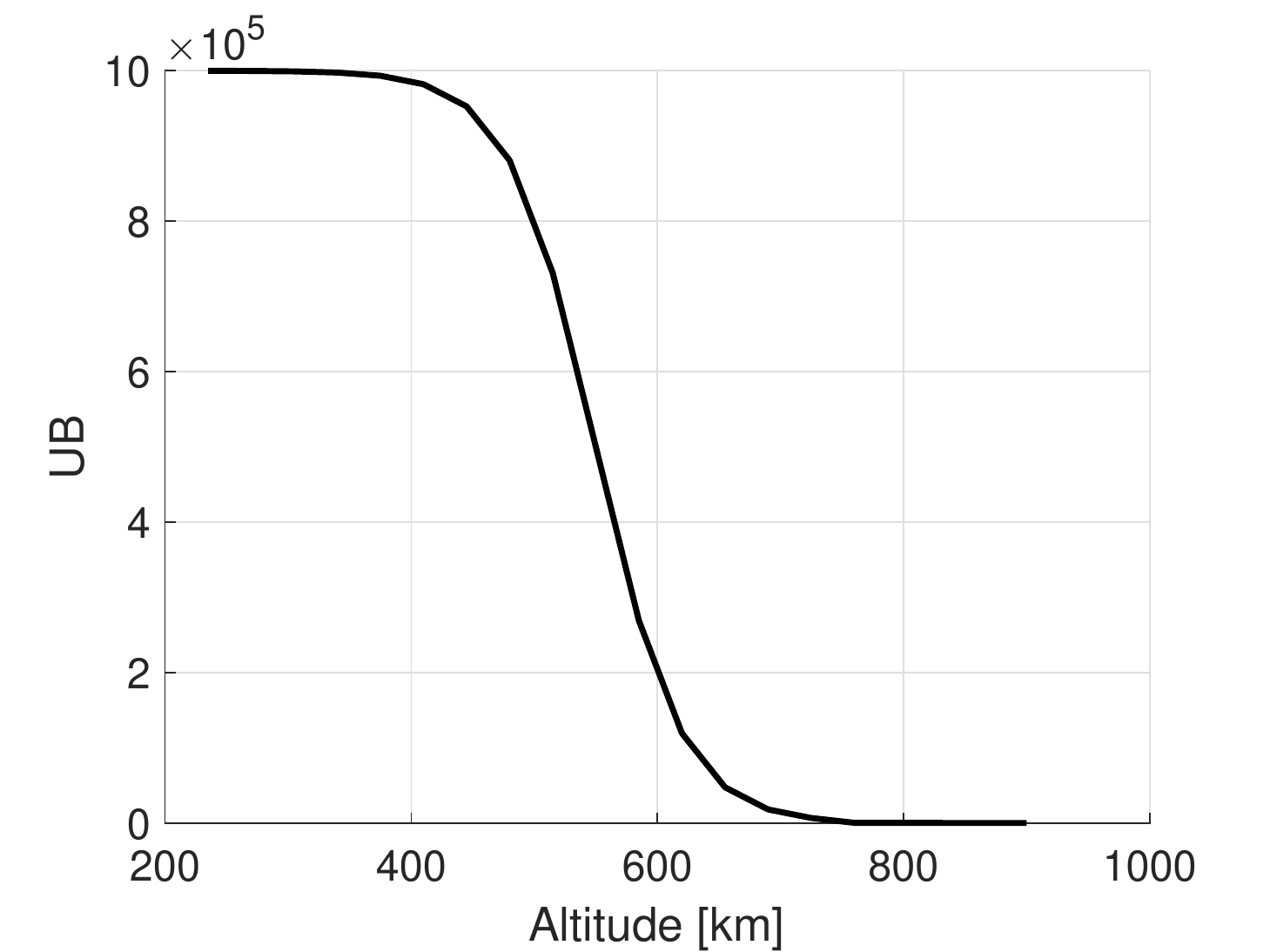}
    \caption{Example of the upper bound representation for $UB_u=10^6$, $UB_l=70$, $h_c=550$ km, $\rho_{UB}=70$ km.}
    \label{fig:UB_tanh}
\end{figure}

In order to guarantee that the optimal launch rate provides a minimum number of satellites launched for each shell, the lower bound ($\textbf{LB}$) of the optimization variables has been set to 10, otherwise the optimizer could have also produced a solution with 0 satellites for a shell, which is not desirable and realistic. Particular attention should be given to the upper bound ($\textbf{UB}$) of the optimization variables. Since we are interested in obtaining the maximum orbital capacity, directly connected to a high launch rate, the upper bound should be set equal to a high number. However, if the same high number is chosen for all the shells, the optimizer is not able to compute an accurate and feasible solution. This is caused by the fact that the optimizer tries to fill the higher shells with many satellites that are going to congest the shells below with many debris and derelicts (and thus collisions) making those shells inaccessible for new launches. To overcome this issue and help the algorithm to converge to a feasible optimal solution, $\textbf{UB}$ has been set as a decreasing function with respect to the altitude. In particular, the values employed for the components of the vector $\textbf{UB}$ have been obtained from the following function:
\begin{equation}
    \label{eq:UB_tanh}
    UB(h) = \frac{1}{2} \left[ (UB_u + UB_l) + (UB_u - UB_l) \tanh \Big(- \frac{h-h_c}{\rho_{UB}} \Big) \right]
\end{equation}
where $UB_u$ and $UB_l$ represent the desired maximum and minimum value of $\textbf{UB}$, $h$ and $h_c$ are the generic altitude and the location of the inflection point of the tanh function, and $\rho_{UB}$ is a constant defined by the user which affects the slope of the curve. A representation of the function in Eq. \eqref{eq:UB_tanh} is provided in Fig. \ref{fig:UB_tanh}. The reason why this function has been chosen is that it is more likely that many satellites can be launched in low altitudes, where the natural decay effects are significant, whereas fewer satellites can be launched in high altitudes, where the atmospheric density is very low and the natural decay does not help to remove satellites, thus obtaining a stable environment over time. In fact, it is worth to observe that the optimizer tends to return an optimal value which is near the maximum of $\textbf{UB}$ for low shells, and near the minimum for high shells, as can be expected. It has been seen from the run of several simulations that increasing $UB_u$ generally leads to obtain a higher capacity but with most of the satellites placed in the lower shells.

\subsection{Optimal Solution and Stability Analysis}

The results of the maximum orbital capacity considering a failure rate of $\bar{\chi}=7\%$ are here reported. In order to start the optimization, the following parameters for Eq. \eqref{eq:UB_tanh} have been chosen: $UB_u=10^6$, $UB_l=70$, $h_c=550$ km, $\rho_{UB}=70$ km. Even considering the smooth function for the upper bound, the algorithm still encountered the problem of filling the higher shells with too many satellites causing the congestion of the shells below, which could not satisfy the failure rate constraint of Eq. \eqref{eq:chi_bar}. For this reason, the upper bound of the launch rate for the last 5 shells have been manually set equal to [330, 150, 130, 100, 70] with a trial-and-error procedure. 

\begin{figure}[!h]
    \centering
    \includegraphics[width=0.8\linewidth]{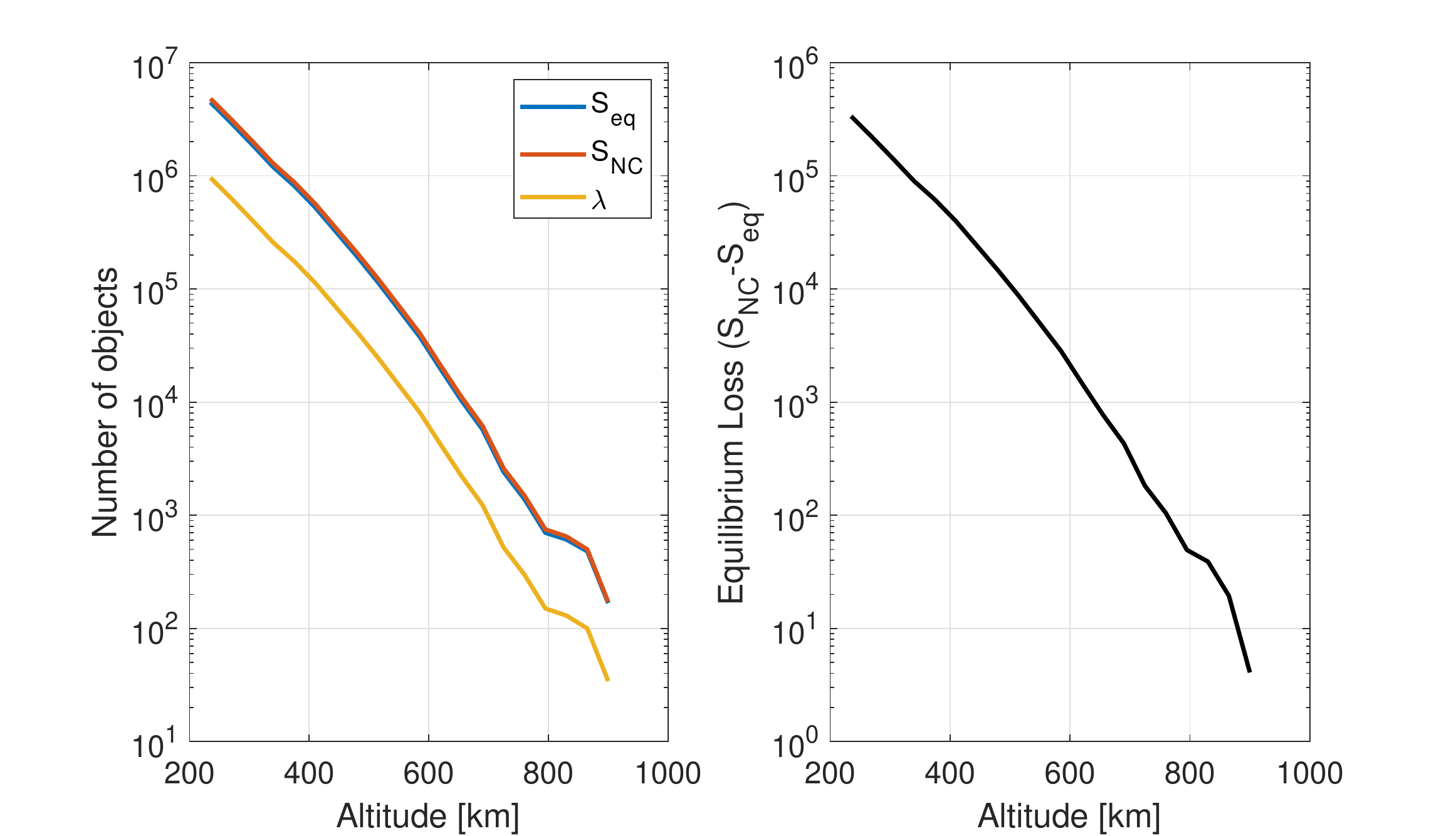}
    \caption{Optimal launch rate and equilibrium solution for active satellites.}
    \label{fig:equilibrium_cap7}
\end{figure}

The computed optimal solution is shown in Fig. \ref{fig:equilibrium_cap7}, where the optimal launch rate and the corresponding equilibrium solution ($S_{eq}$), in terms of active satellites, are reported in the left plot as functions of the altitude. Please note that a logarithmic scale is adopted on the y-axis to better visualize the results. In particular, it is evident that the blue curve and the red curve, associated to the ideal equilibrium solution ($S_{NC}$), appears almost overlapped. Indeed, along the curves, the difference does not exceed the 7 \% which represents the failure rate constraint. The difference $(S_{NC}-S_{eq})$, here called equilibrium loss, is illustrated in the right plot of Fig. \ref{fig:equilibrium_cap7}, from which we can observe that higher deviations from the ideal solution occur in lower altitude shells, where more satellites can be fit and thus more collisions are present. However, in lower shell, the stronger atmospheric density plays a key role in removing derelicts and debris. The maximum capacity obtained from this optimal solution is about 12.6 millions of satellites (see also Table \ref{tab:fail_rate_diff}), with a launch rate of 2.7 millions of satellites per year.

\begin{figure}[!h]
    \centering
    \includegraphics[width=0.8\linewidth]{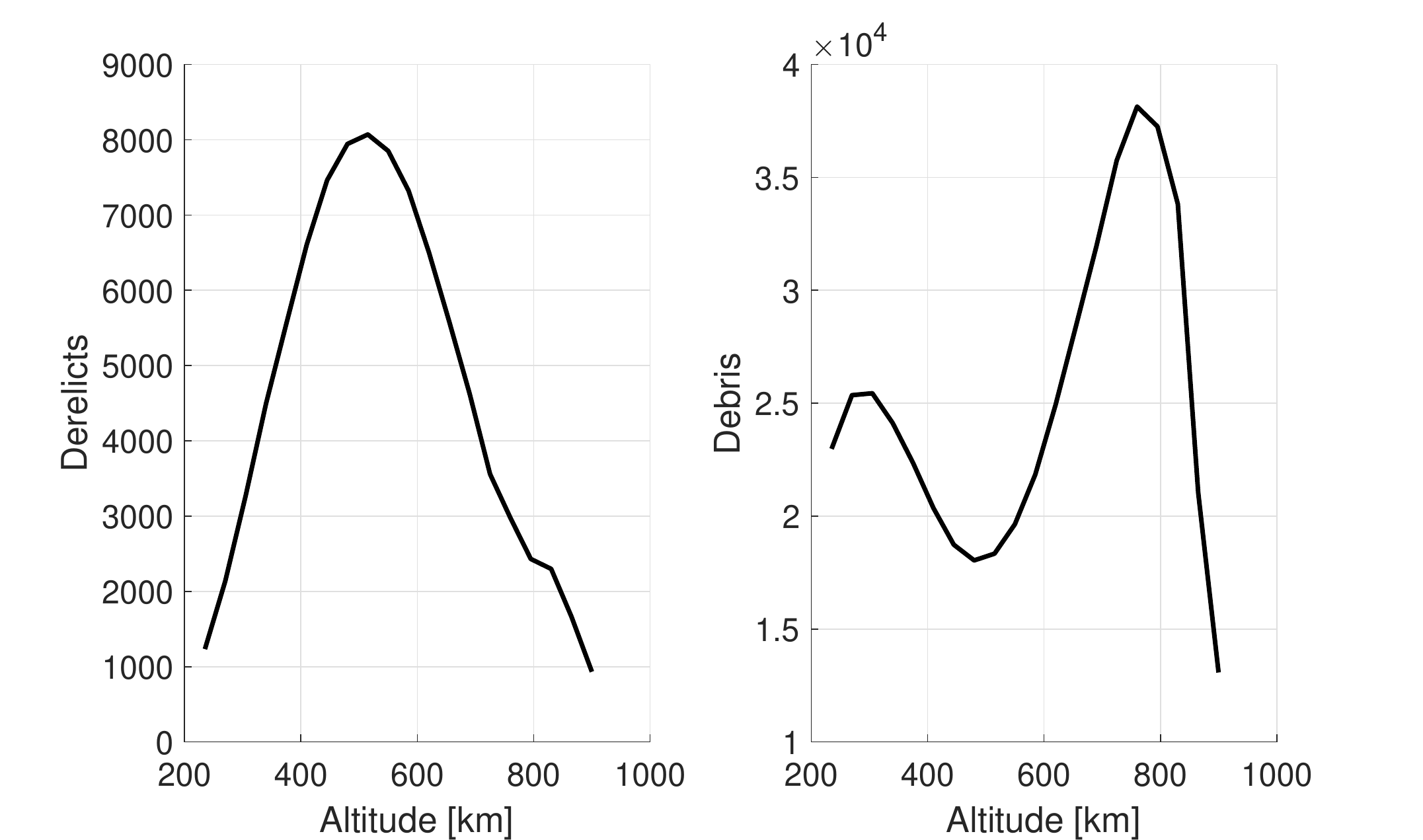}
    \caption{Optimal equilibrium solution for derelicts and debris.}
    \label{fig:D_N_cap7}
\end{figure}

The distribution of derelicts and debris for the computed optimal equilibrium solution is shown in Fig. \ref{fig:D_N_cap7}. The trend of the derelicts indicates that the peak is around 500 km of altitude, where enough satellites can be fit but the atmospheric drag is not able to efficiently compensate with the orbital decay. At the same time, a low number of derelicts is present in very low and high altitude shells. For the first ones, millions of satellites can be launched but the derelicts are rapidly disposed by the atmospheric drag. For the high shells, a low number of derelicts is present since a low number of satellites can be launched and the atmospheric drag is not efficient. On the other hand, the trend of the debris (right plot of Fig. \ref{fig:D_N_cap7}) present one higher peak at high altitudes, where again the atmospheric density is very low, and another one at low altitudes where many satellites can be placed. One last consideration can be carried out for the aforementioned figures. The less smooth behaviour occurring around 800 km of altitude for both the active satellites and the derelicts is due to the chosen values of the upper bound for the launch rate. It is worth mentioning again that without using those boundaries the upper shells would have been fully filled, with the consequent result of overly congesting the shells below, leading to the underestimation of the overall maximum capacity.

\begin{figure}[!h]
    \centering
    \includegraphics[width=0.8\linewidth]{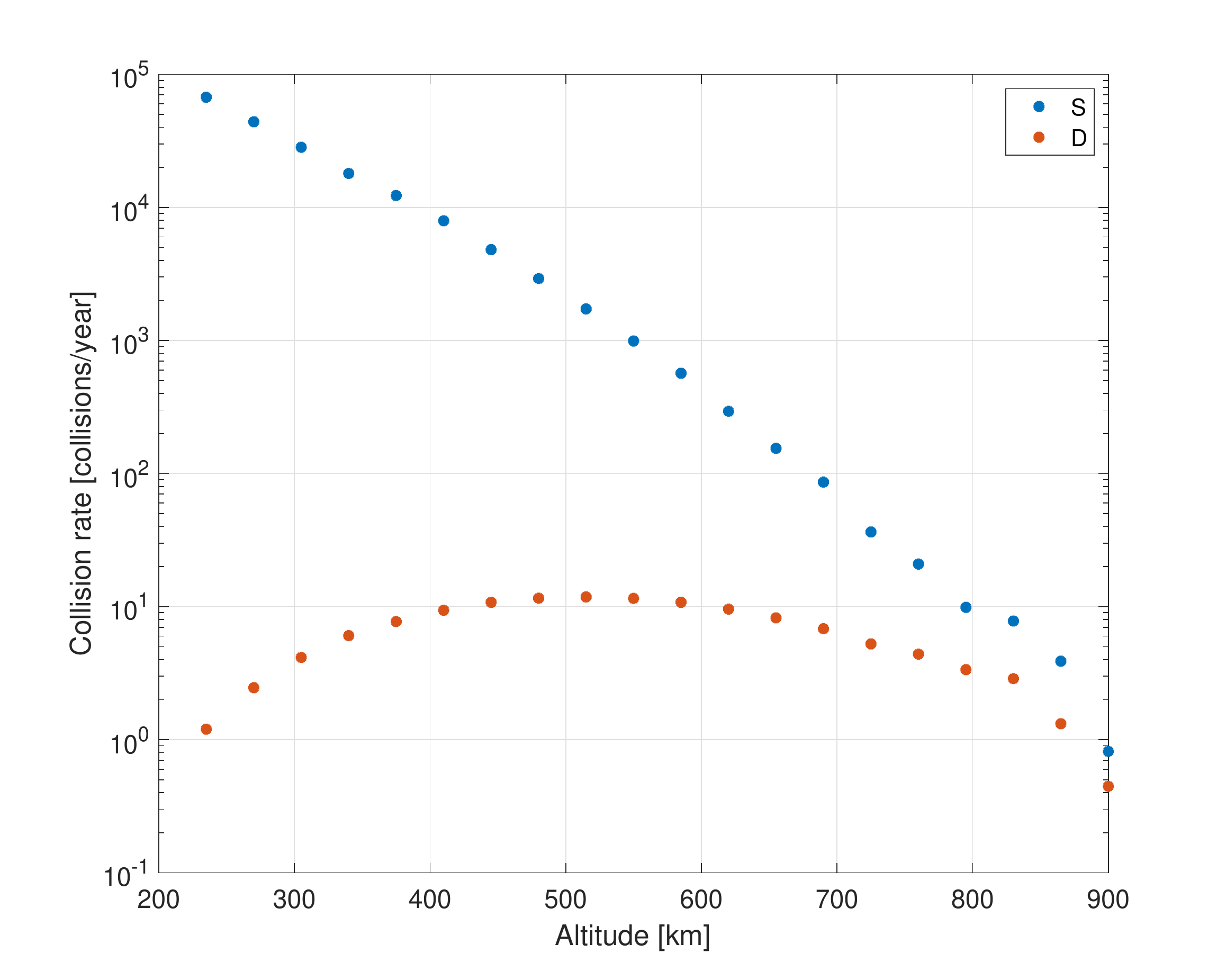}
    \caption{Collision rate for the optimal solution.}
    \label{fig:collision_rate_cap7}
\end{figure}

Figure \ref{fig:collision_rate_cap7} illustrates the collision rate for the active satellites and derelicts as a function of the altitude. As can be seen, the trends look similar to the trends of the curves $S_{eq}(h)$ and $D_{eq}(h)$, shown in Figs. \ref{fig:equilibrium_cap7} and \ref{fig:D_N_cap7} respectively. This is reasonable since more collisions occur in the shells where more objects are located. For example, analyzing the lowest shell, we can see that the number of collisions per year is about 50,000, which represents 1.5\% of the total number of satellites in that shell (about 3.5 million). This high collision rate is strongly related to the chosen failure rate and the high number of satellites in that shell. While this collision rate is very high, it highlights an important aspect of the modeling approach. The chosen equilibrium condition will prevent secular growth in species populations, but does not enforce any constraints relating to short-term safety of flight. Thus, while lower shells may not lead to run-away growth in debris populations due to the strength of atmosphere drag effects, they still may be sufficiently hazardous to pose high or unacceptable cost and threat to safety of flight/mission assurance for operators that might make use of them. To address these short-term safety of flight and mission assurance questions, further constraints could be added to system optimization relating to acceptable collision rates (either on an aggregate or per-species-pair basis). In reality, it is also potentially unlikely that there is a short to medium term business case that would saturate these lower shells with that number of satellites shown in the model.

\begin{figure}[!h]
    \centering
    \includegraphics[width=0.7\linewidth]{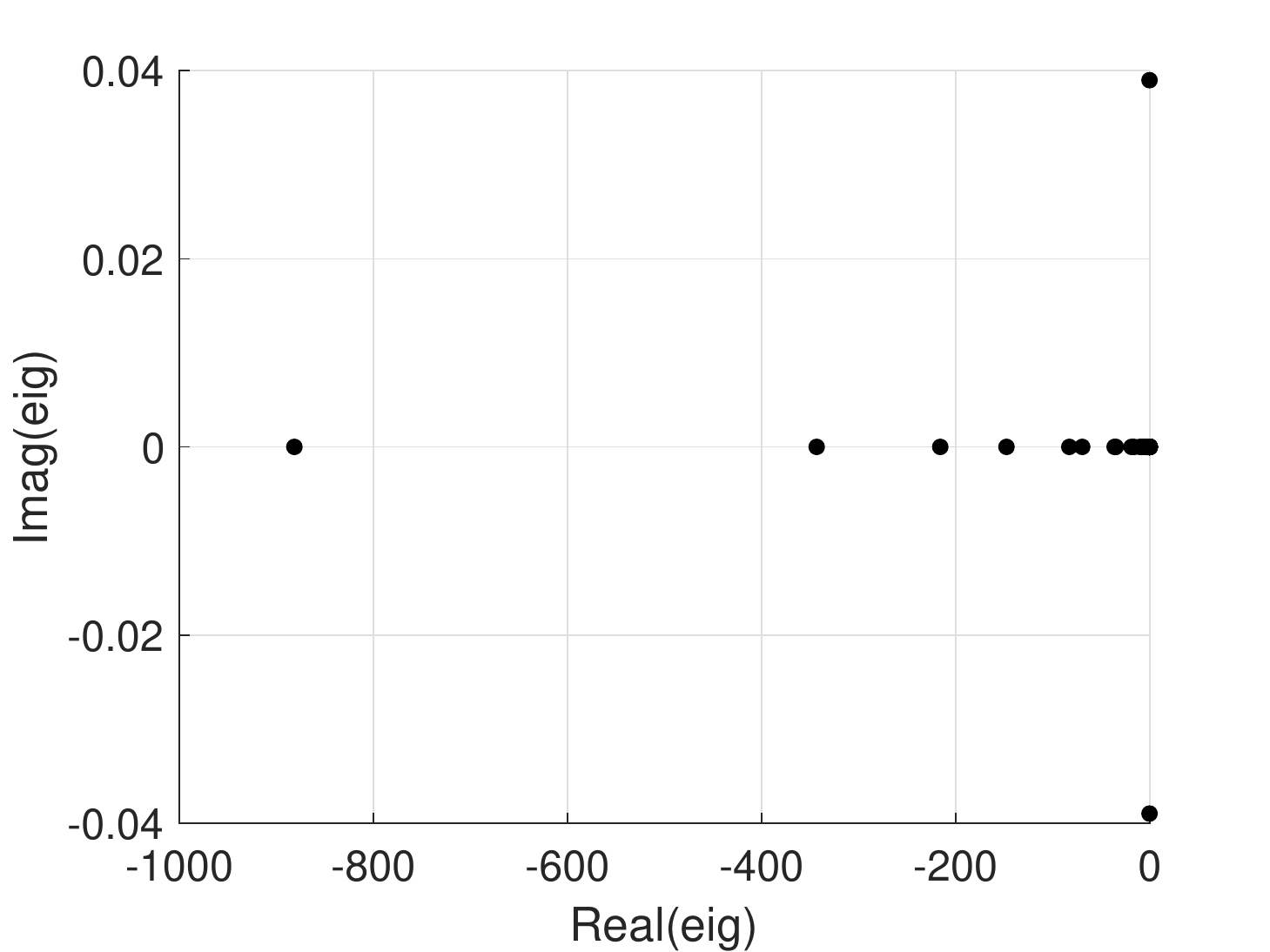}
    \caption{Eigenvalues associated to the optimal solution.}
    \label{fig:eig_cap7}
\end{figure}

\begin{figure}[!h]
    \centering
    \includegraphics[width=0.7\linewidth]{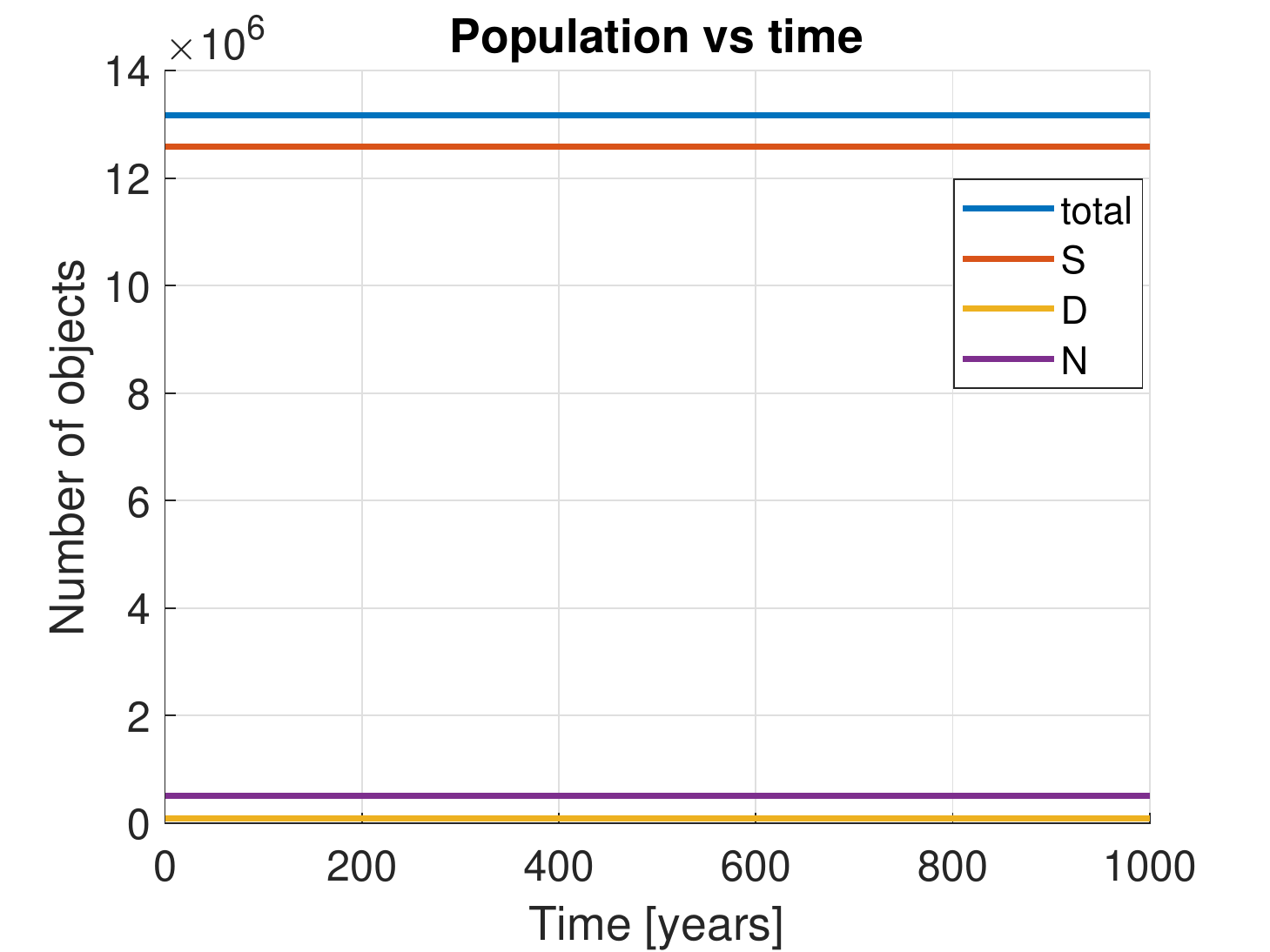}
    \caption{Propagation of the MOCAT-3 model as a function of time.}
    \label{fig:integration_cap7}
\end{figure}

\begin{figure}[!h]
    \centering
    \includegraphics[width=0.7\linewidth]{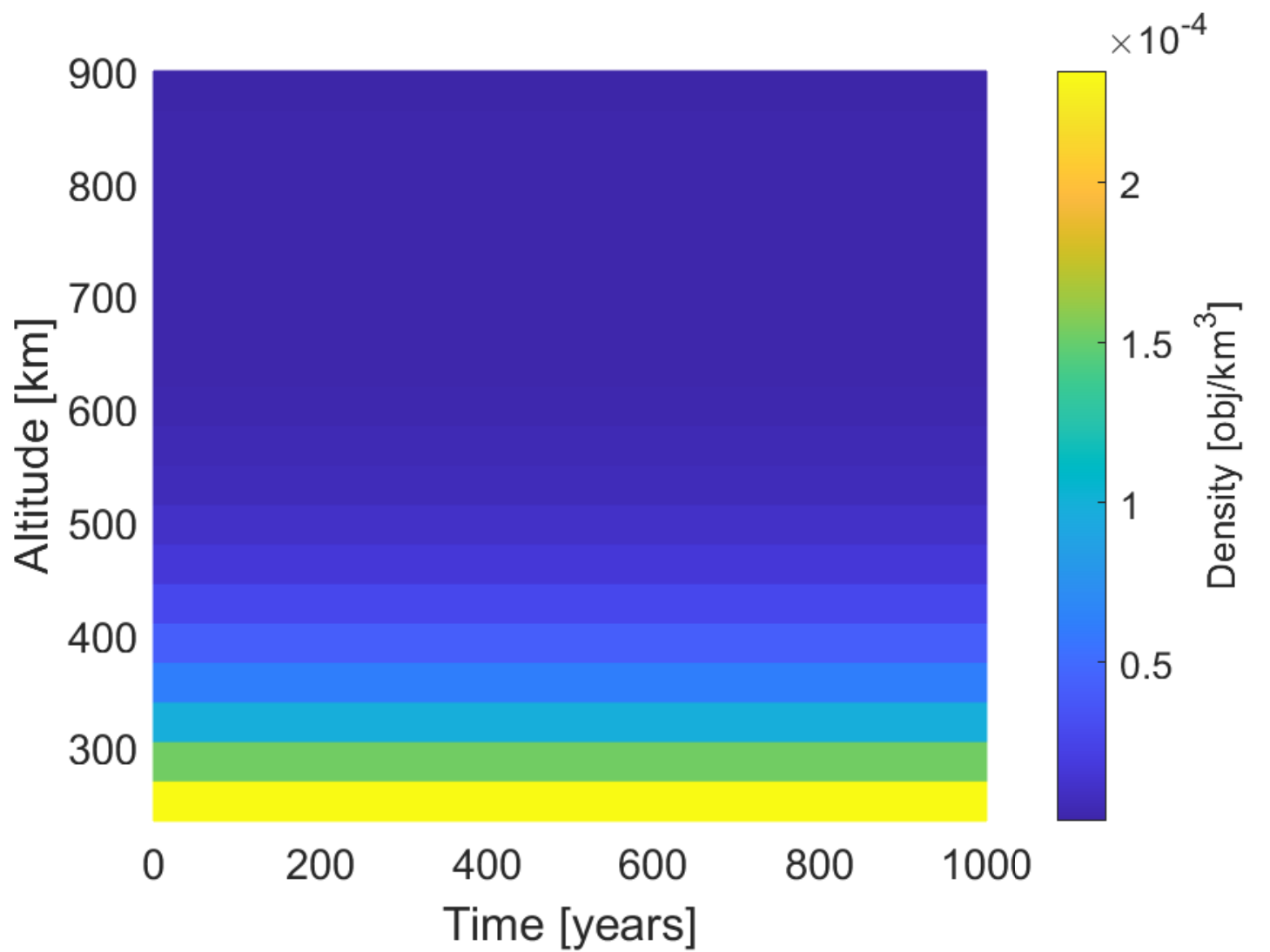}
    \caption{Density of the objects as a function of time and altitude for the optimal solution.}
    \label{fig:density_cap7}
\end{figure}

In order to test the stability of the equilibrium solution, the eigenvalues associated to the jacobian matrix of the system have been computed. Figure \ref{fig:eig_cap7} reports all the eigenvalues: it is possible to notice that all of them are negative real number, apart from a pair of complex conjugate eigenvalues, which however have a negative real part. This proves the stability of the system with a little oscillatory behaviour of the solution, represented by the small imaginary part of the two complex eigenvalues. 
The stability of the solution is also proved by Fig. \ref{fig:integration_cap7}, where each species of ASOs is represented as a function of time, showing a constant trend. Finally, the density of the objects in the analyzed region of LEO is illustrated in Fig. \ref{fig:density_cap7} as a function of both time and altitude, where it is possible to see that the most crowded altitudes are the lower shells. 

\begin{figure}[!h]
    \centering
    \includegraphics[width=0.7\linewidth]{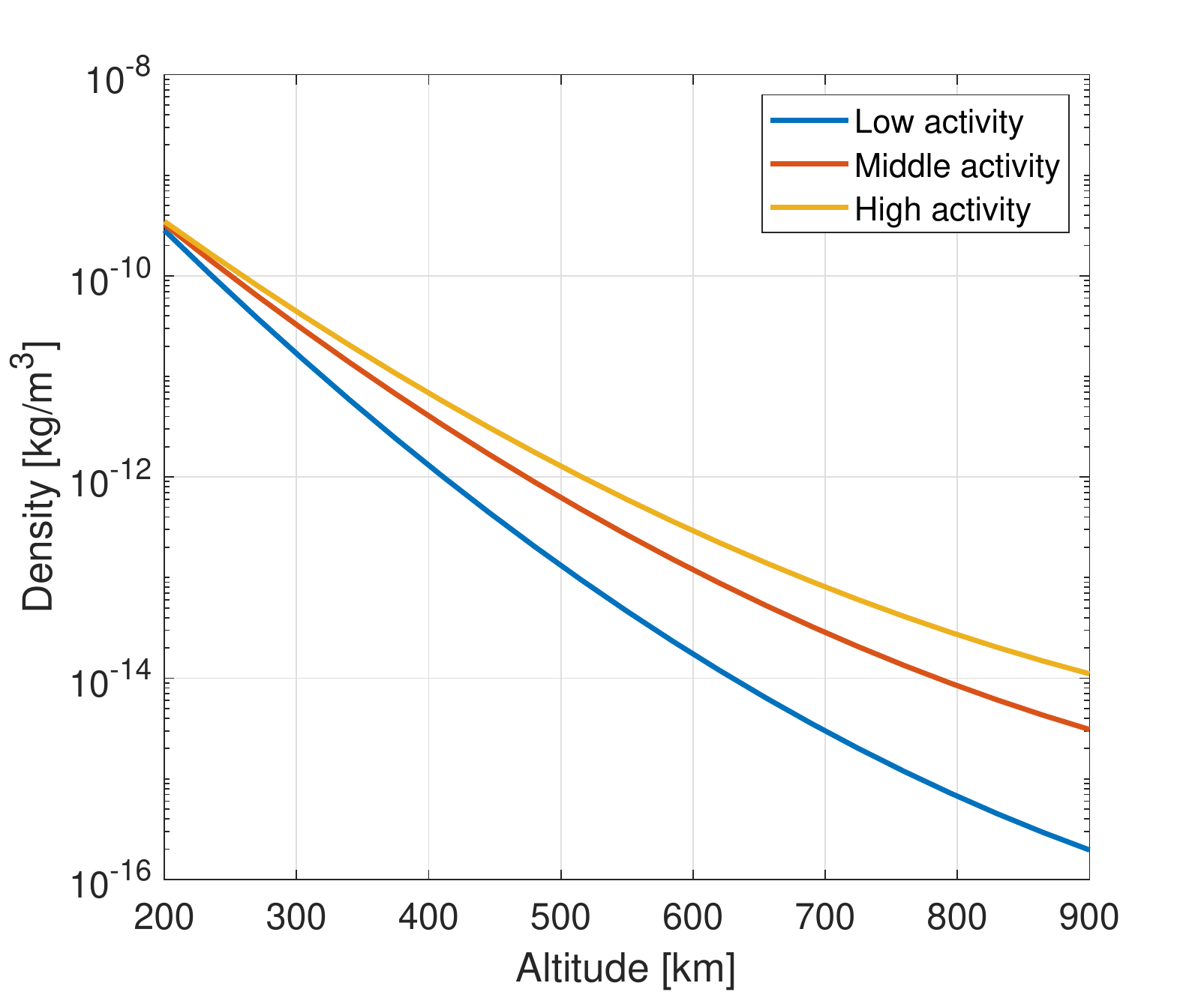}
    \caption{Density profiles for low, middle and high solar and geomagnetic activity.}
    \label{fig:density_profiles}
\end{figure}

\begin{figure}[!h]
    \centering
    \includegraphics[width=0.8\linewidth]{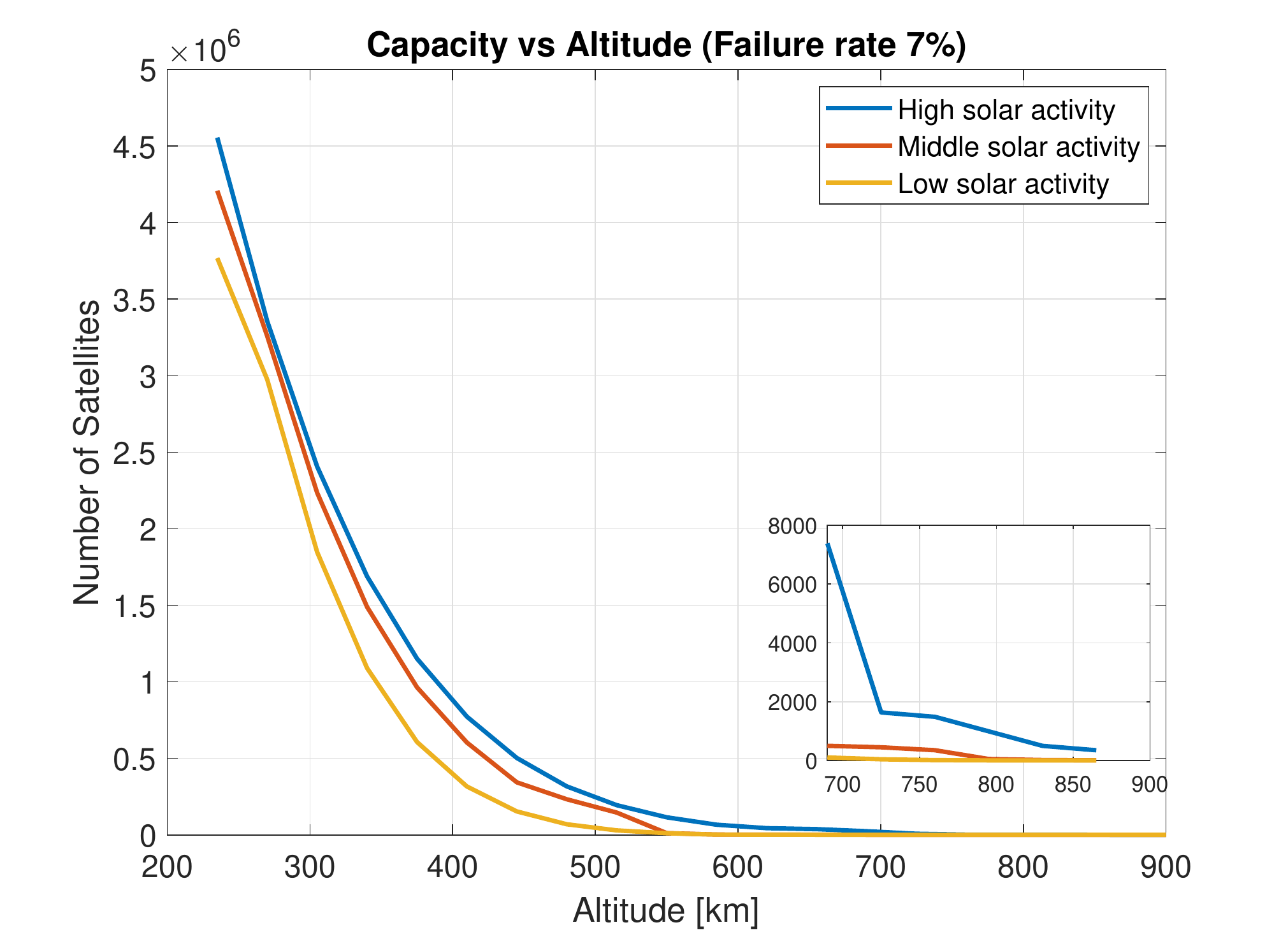}
    \caption{Comparison of the capacity for different density profiles.}
    \label{fig:capacity_altitude_cap7_density}
\end{figure}

To address what are the consequences of the space weather, in terms of solar and geomagnetic activities, on the orbital capacity, the static exponential density model can be replaced by the following density model \cite{acedo2017kinematics}: 
\begin{equation}
    \label{eq:density_solar}
    \begin{cases}
    T = 900 +2.5(F_{10.7}-70)+1.5A_p \\
    m = 27-0.012(h-200) \\
    H = T/m \\
    \rho = 6\cdot 10^{-10} \exp(-(h-175)/H)
    \end{cases}
\end{equation}
where $F_{10.7}$ is the solar radio flux, measured in solar flux units (SFUs) with one SFU equal to $10^{-22}$ W/m$^2$Hz, and $A_p$ is the geomagnetic index. Indeed, in periods of strong solar and geomagnetic activity, both the values of $F_{10.7}$ and $A_p$ increase, heating up the atmosphere and increasing the atmospheric density. Three scenarios are taken into account: one concerning high activities ($F_{10.7}=200$ and $A_p=8$), one with middle activities ($F_{10.7}=150$ and $A_p=5$), and one with low activities ($F_{10.7}=70$ and $A_p=2$). The corresponding density profiles of the analyzed space weather conditions are illustrated in Fig. \ref{fig:density_profiles}. The results of the optimization for the three scenario, always considering a 7\% of failure rate, are reported in Table \ref{tab:fail_rate_diff} and represented in Fig. \ref{fig:capacity_altitude_cap7_density} in terms of capacity vs altitude. As expected, more satellites can be fit into orbit when high solar and geomagnetic activities are considered, which lead to an overall greater density and thus a stronger drag sinking mechanism.

\subsection{Comparison for Different Failure Rates}

\begin{figure}[!h]
    \centering
    \includegraphics[width=0.9\linewidth]{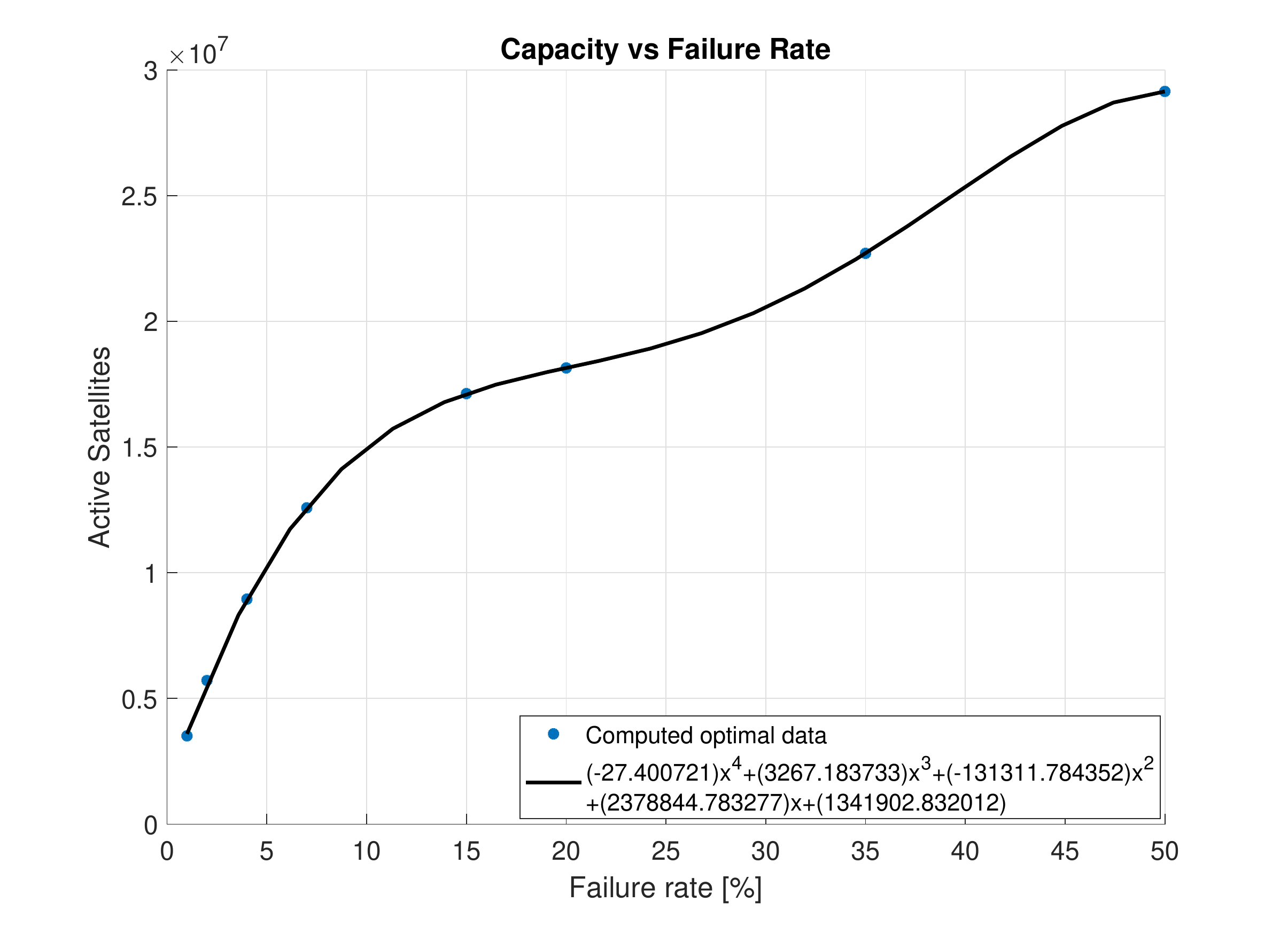}
    \caption{Capacity as a function of the failure rate.}
    \label{fig:failure_rate_capacity}
\end{figure}

In order to test how the capacity changes according to the variation of the failure rate $\bar{\chi}$, different optimization procedures have been run by varying $\bar{\chi}$. The results are shown in Table \ref{tab:fail_rate_diff} and Fig. \ref{fig:failure_rate_capacity}. As can be seen, greater failure rates allow to obtain a higher capacity with more active satellites. However, in these cases, the number of debris also increases. According to the ESA's annual space environment report\footnote{\url{https://www.esa.int/Safety_Security/Space_Debris/Space_debris_by_the_numbers}}, the number of debris regularly tracked by Space Surveillance Networks and maintained in their catalogue is about 31,300, whereas the number of debris estimated, based on statistical models (MASTER-8), to be in orbit is 36,500 for space debris objects greater than 10 cm, 1 million for space debris objects from 1 cm to 10 cm, and 130 million for space debris objects from 1 mm to 1 cm. From the results shown in Table \ref{tab:fail_rate_diff}, we can see that the number of debris found in the proposed solutions is generally greater than the value reported in the ESA's annual report for debris with size greater than 10 cm. 
Moreover, the computed solutions indicate that, within this model, the LEO environment remains sustainable even with more debris objects than exist today (at least in these altitudes and species distribution). Finally, for the range of analyzed failure rates [1,50] \%, a good approximation of the maximum orbital capacity as a function of the failure rate (i.e., $S_{eq,max}(\bar{\chi})$) is provided by the following polynomial expression:
\begin{equation}
    \label{eq:cap_chi_fitting}
    S_{eq,max}(\bar{\chi}) = -27.40 \bar{\chi}^4 + 3,267.18 \bar{\chi}^3-131,311.78 \bar{\chi}^2+ 2,378,844.78 \bar{\chi} +1,341,902.83
\end{equation}

\begin{table}[htb!]
\small
    \centering
    \begin{tabular}{ccc}
    \hline
    Failure Rate [\%] & Active Satellites & Debris\\
    \hline
    1 & $3.5\cdot10^6$ & $3.4\cdot10^4$  \\
    2 & $5.7\cdot10^6$ & $9.2\cdot10^4$ \\
    4 & $8.9\cdot10^6$ & $2.7\cdot10^5$ \\
    7 (static) & $12.6\cdot10^6$ & $5.0\cdot10^5$ \\
    7 (low activity) & $10.8\cdot10^6$ & $3.75\cdot10^5$ \\
    7 (middle activity) & $13.5\cdot10^6$ & $1.70\cdot10^5$ \\
    7 (high activity) & $15.2\cdot10^6$ & $2.93\cdot10^5$\\
    20 & $18.1\cdot10^6$ & $8.5\cdot10^5$ \\
    35 & $22.7\cdot10^6$ & $2.37\cdot10^6$  \\
    50 & $29.1\cdot10^6$  & $4.08\cdot10^6$  \\
    \hline    
    \end{tabular}
    \caption{Optimal results for different failure rates.}
    \label{tab:fail_rate_diff}
\end{table}

\subsection{Accommodating for the Future Launch Plans}

\begin{figure}[!h]
    \centering
    \includegraphics[width=1.\linewidth]{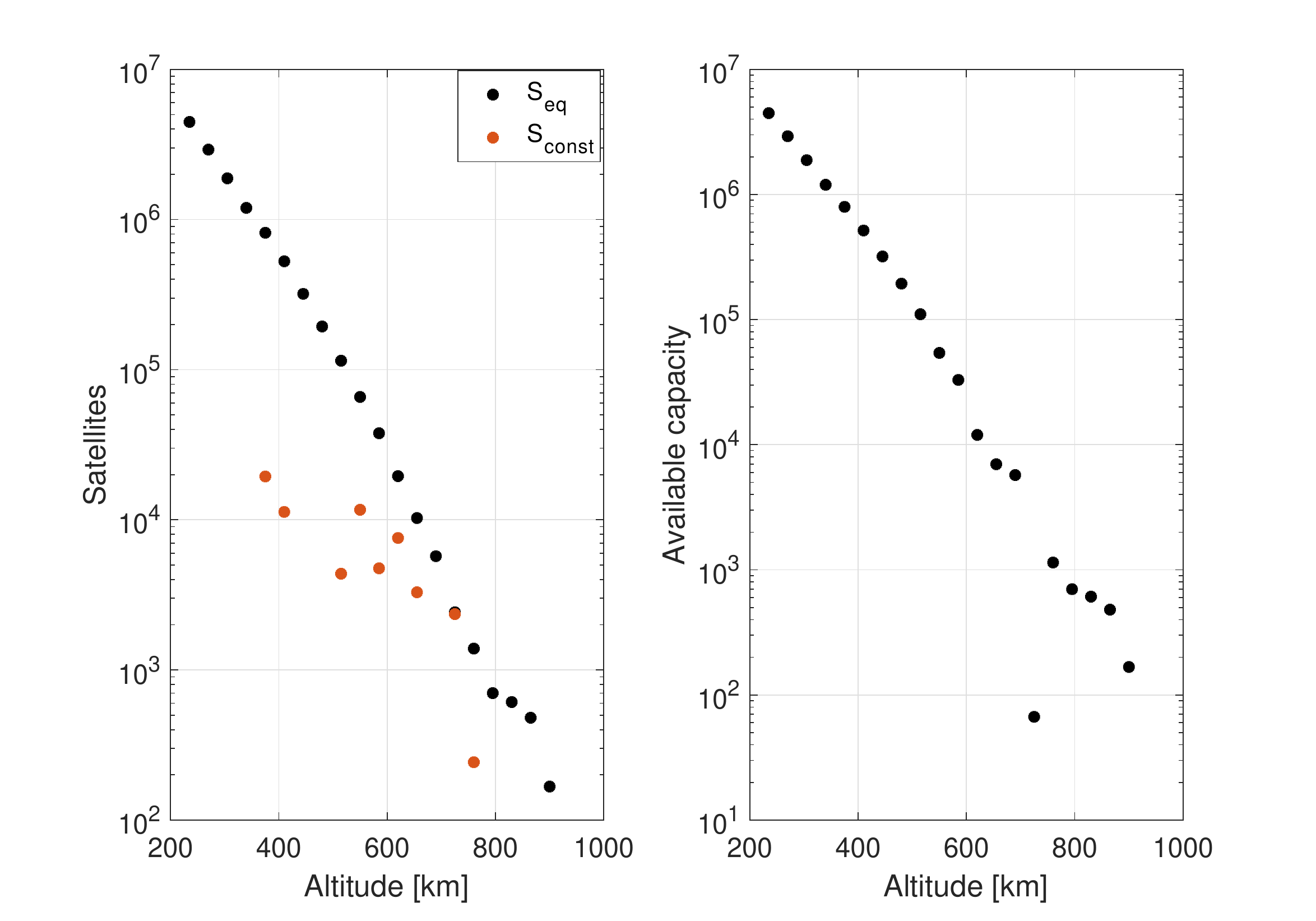}
    \caption{Analysis of the capacity according to the future planned constellations (the coefficients employed in the MOCAT-3 are taken from the JASON report \cite{jason}).}
    \label{fig:comparison_constellations_cap7}
\end{figure}

Another interesting analysis consists of comparing the obtained results with demand data for selected future planned satellite constellations within the [200,900] km altitude range, reported in Table \ref{tab:future_const} based on regulatory filings. This is useful to understand if the obtained solutions could accommodate planned future launch traffic. This future traffic was binned into altitude shells to generate a launch demand curve as a function of altitude. As can be observed in the left plot of Fig. \ref{fig:comparison_constellations_cap7} obtained for 7\% of failure rate and a static exponential density model, the number of active satellites of the proposed optimal solution ($S_{eq}$) is always greater than the number requested by this particular set of constellations ($S_{const}$). This means that the obtained solution could be compliant with the future traffic launches and that there could still be available capacity (right plot of Fig. \ref{fig:comparison_constellations_cap7}) that can be exploited while maintaining at the same time a sustainable and stable LEO environment. Furthermore, all the analyzed scenarios reported in Table \ref{tab:future_const} would actually allow to accommodate for the future traffic launches. This result demonstrates the type of analysis that is possible to carry out with this modeling approach, but it is important to note that, until this model is verified against other models and empirical data, it is uncertain whether this conclusion would hold true for the behavior of the actual space environment.

\begin{figure}[!h]
    \centering
    \includegraphics[width=0.9\linewidth]{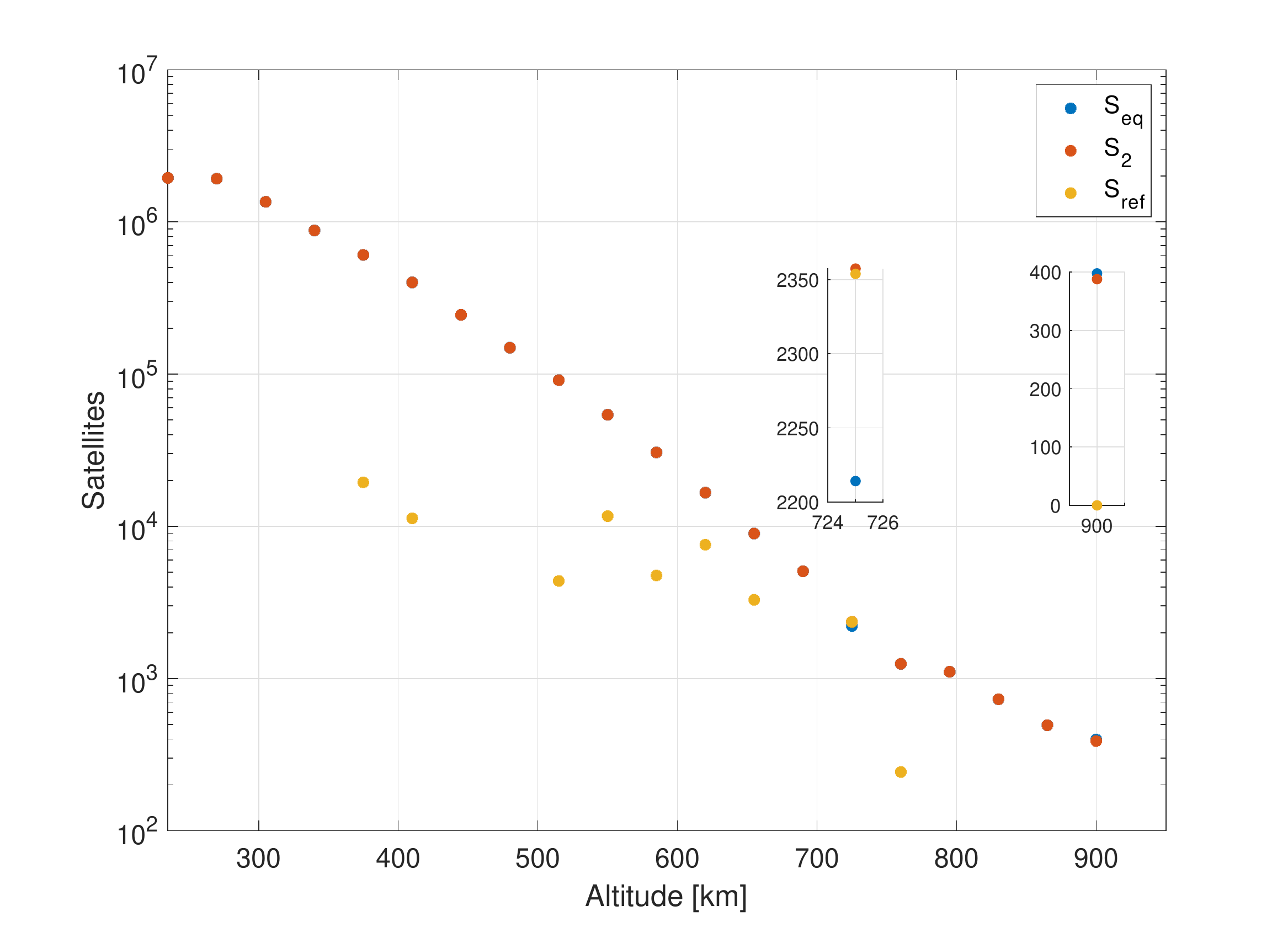}
    \caption{Results of the accommodation procedure for case A (the coefficients employed in the MOCAT-3 are taken from the JASON report \cite{jason}).}
    \label{fig:launc_accom_cap7_sup}
\end{figure}

It can also be useful to consider a case for which the future traffic launches demand is not satisfied by the computed optimal solution. For this case, an approach to still try to accommodate the future needs is here proposed. The main idea is that it is possible to decrease the launch rate of some shells (either above or below the one that does not accommodate for the future plans) and increase the launch rate in the shell where it is required. This is carried out at the expense of the maximum capacity or the minimum number of launched satellites that should be guaranteed per shell. Two different examples are provided below, considering a solution with 7\% of failure rate that in principle does not accommodate for the future satellite demand in one orbital shell around 725 km of altitude. In fact, the number of satellites requested by the future demand for that altitude shell is 2354 (see Table \ref{tab:future_const}). However, the proposed solution can only fit 2214 satellites in that shell (see Fig. \ref{fig:launc_accom_cap7_sup} where the computed blue dot is lower than the reference yellow dot). Hence in the first case, denoted as case A, the proposed strategy is to increase the launch rate of the shell that we are interested in of 30 satellites per year more, and decrease at the same time the launch rate of the highest shell of just 2 satellites per year (no satellites are required in the highest shell around 900 km according to Table \ref{tab:future_const}). The effects of this change can be visualized by propagating the MOCAT-3 model, considering the equilibrium solution and the new launch rates, for 1000 years. The results at the end of the propagation are provided in Fig. \ref{fig:launc_accom_cap7_sup} in terms of the capacity as a function of the altitude. It is possible to see that the new solution ($S_2$) with the modified launch rate is able to accommodate the future needs, allowing 2356 satellites in that shell, thus satisfying the requirement (see Fig. \ref{fig:launc_accom_cap7_sup} where the computed red dot is higher than the reference yellow dot). The reason why this is possible is that less derelicts and debris flow into lower shells, with consequent less collisions, allowing to launch more satellites in the considered shell. Thus, for this case A, the accommodation of the reference solution is obtained at the expense of a lower number of satellites that can be launched in an upper shell. At this point, an important consideration must be done. Changing the launch rate actually changes the equilibrium points of the system. However, the feasibility of the new accommodated solution can be checked by looking at the total number of debris at the end of the propagation period ($N_2$). If it is lower or equal than the total number of debris of the reference optimal solution ($N_{eq}$), then the new solution is considered feasible, since the space environment doesn't change too much and the new launch rate doesn't cause a growth of the space debris population, which would be undesired. Indeed, for case A, the number of debris at final time is lower than the optimal solution, in particular $(N_2-N_{eq})\sim-1154$.

\begin{figure}[!h]
    \centering
    \includegraphics[width=0.9\linewidth]{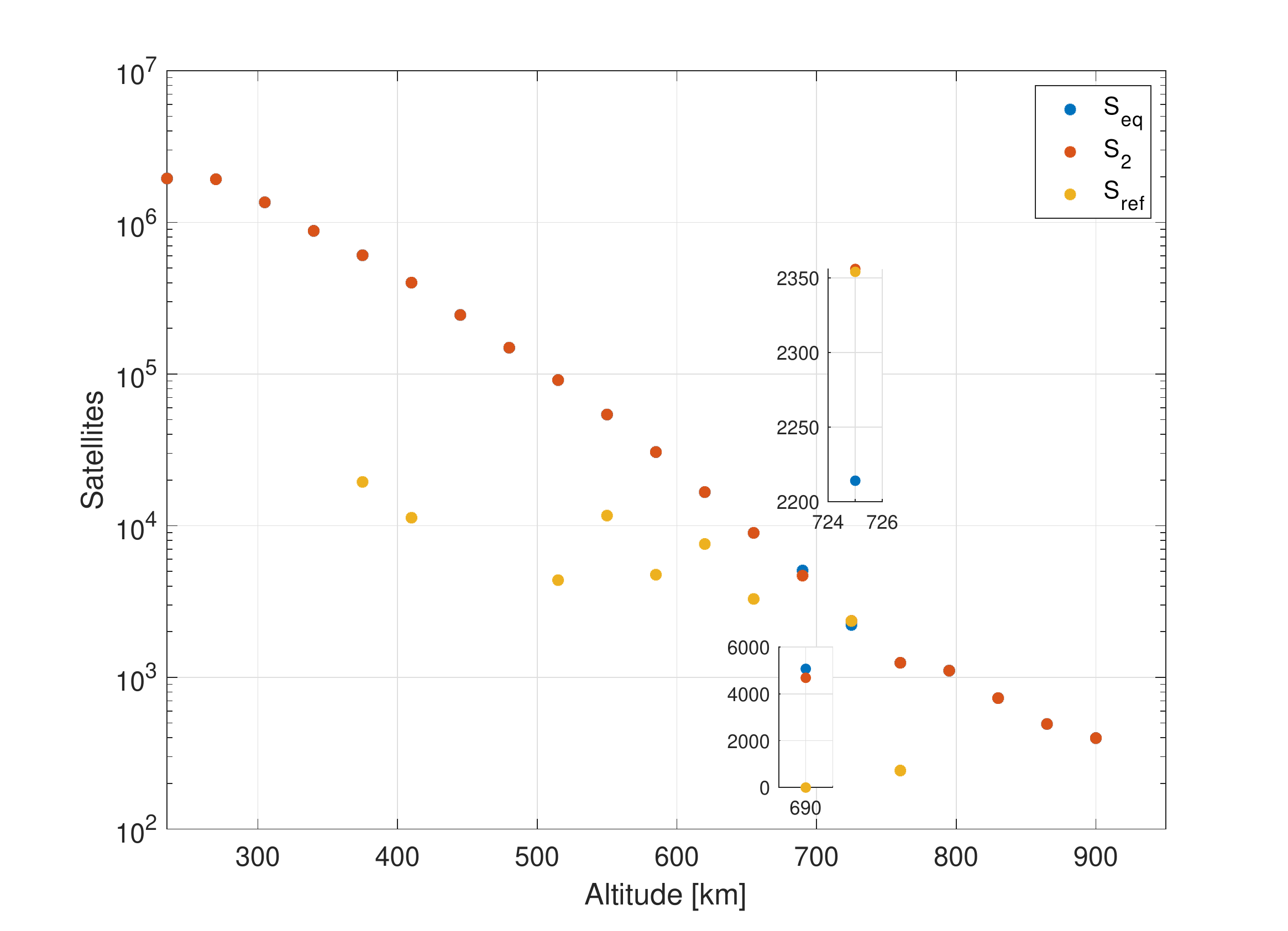}
    \caption{Results of the accommodation procedure for case B (the coefficients employed in the MOCAT-3 are taken from the JASON report \cite{jason}).}
    \label{fig:launc_accom_cap7_inf}
\end{figure}

For case B, the launch rate of the shell that we are interested in is still increased by 30 satellites per year, but the launch rate is now decreased in a shell below. In fact, the launch rate of the shell right below the considered one is decreased of 80 satellites per year. This actually allows to increase the launch rate, and thus the number of objects, in the considered shell allowing for more objects to flow in the shell below, whose total number of objects is compensated by the reduction of the corresponding launch rate. The results obtained with the strategy of case B are reported in Fig. \ref{fig:launc_accom_cap7_inf}, where the same considerations of the previous example are still valid. For this case B, $(N_2-N_{eq})\sim-290$, which proves the feasibility of the strategy.

\begin{table}[htb!]
\small
    \centering
    \begin{tabular}{cccc}
    \hline
    Name & Altitude [km] & Number of satellites & Inclination [deg]\\
    \hline
AST SpaceMobile	& 737,5	 & 75	& 55 \\
AST SpaceMobile	& 732,5	 & 150	& 40 \\
AST SpaceMobile	& 727,5	 & 18	& 0 \\
Inmarsat	    &  724	 & 18	& 0 \\
Astra	        &  700	 & 1792	& 55 \\
Astra	        &  700	 & 40	& 0 \\
Astra	        &  690	 & 504	& 98 \\
Amazon (Kuiper-V) &	650 &	325	  & 80 \\
Amazon (Kuiper-V)&	640 &	652	  &  72 \\
Amazon (Kuiper-Ka)&	630 &	1156  & 51,9 \\
Amazon (Kuiper-V)&	630 &	1156  & 51,9 \\
SpaceX Gen2 C1&	614	&324 &	115,7  \\
Amazon (Kuiper-Ka)&	610 &	1296  &    42 \\
Amazon (Kuiper-V)	&610 &	1296  & 	42 \\
SpaceX Gen2 C1	&604&	144 &	148 \\
CASC China SatNet (Guowang)&	600&	2000&	50 \\
Kepler Communications	&600&	360&	89,5 \\
Orora.Tech	&600&	100&	97,6 \\
Amazon (Kuiper-Ka)&	590&	784	&33 \\
Amazon (Kuiper-V)&	590	&784	&33 \\
CASC China SatNet (Guowang)	&590	&480	&85 \\
SpaceX Gen1	& 570	&1584	&70 \\
SpaceX Gen1	& 560	&1584	&97,6 \\
SpaceX Gen1	& 550	&1584	&53 \\
SpaceX Gen1	& 540	&1584	&53,2 \\
SpaceX Gen2 C1	&535	&3360	&33 \\
SpaceX Gen2 C1	&530	&3360	&43 \\
SpaceX Gen2 C1	&525	&3360	&53 \\
CASC China SatNet (Guowang)&	508	&3600	&55 \\
Planet (various altitudes, no prop.)	&500	&200&	97,5 \\
Spire Global (various altitudes, no prop.)&	500 &	275	&97,4 \\
Satelllogic (various altitudes, no prop.)	&486,65	&300&	97,4 \\
Astra	&400	&4148	&55 \\
Astra	&390	&4896	&30 \\
Astra	&380	&2240	&97 \\
SpaceX Gen2 C1	&360	&3600	&96,9 \\
SpaceX Gen2 C1	&350	&5280	&38 \\
SpaceX Gen2 C1	&345	&5280	&46 \\
SpaceX Gen2 C1	&340	&5280	&53 \\
    \hline    
    \end{tabular}
    \caption{Data of future planned satellite constellations. Most of these data are collected from the International Bureau Filings of the Federal Communications Commission  (\url{https://fcc.report/IBFS}) and the e-Submission of Satellite Network Filings of the International Telecommunication Union (\url{https://www.itu.int}).}
    \label{tab:future_const}
\end{table}

\section{Conclusions}

This paper focused on the analysis of orbital capacity of Low Earth Orbit carried out through the development of a new source-sink model, called MOCAT-3, which is a multi-bin multi-species model, including active satellites, derelicts and debris. The model simulates the evolution of LEO environment and related anthropogenic space object populations. A new definition of orbital capacity has been proposed, based on dynamical analysis and equilibrium solutions of systems of ODEs, that is related to the long-term stability of LEO. An optimization procedure has been carried out to select the launch rate which provides the highest capacity for a particular assumed failure rate constraint. The proposed procedure has been applied to the orbital altitudes belonging to the range 200-900 km. The results have shown a maximum capacity of about 12.6 millions of active satellites if a failure rate of 7\% is considered, the majority of which are located in lower shells where the atmospheric density and orbit decay effects are higher. The launch rate corresponded to 2.7 millions of satellites launched per year. Moreover, stability analysis for the proposed solution has been reported, proving the long-term sustainability of the studied region of LEO with the computed launch rate. Particularly, the computed solution appears to be asymptotically stable (all the eigenvalues have negative real part) with a oscillatory behaviour around the obtained equilibrium condition (due to a pair of complex eigenvalues). In addition, other analysis taking into account different solar and geomagnetic level of activity have been provided, showing that higher space weather activity leads to a greater capacity, as expected.

In order to demonstrate a method to estimate how capacity varies according to the failure rate, different failure rates have been considered within the range [1,50] \%. Future launch plans have been considered to test the stability of this population against the available capacity obtained in the proposed solution under equilibrium conditions. In case of no compliance in some altitude shells, a strategy based on the variation of the launch rate with respect to the reference optimal solution has been proposed to accommodate future demand. For all these results, it is important to note that validation and verification of MOCAT-3 has yet to be carried out. While the proposed methods are generalizable, it has not yet been proven that model outputs correspond to outcomes from other established modeling approaches or empirical data. The specific results obtained and resulting conclusions could shift after calibration of model coefficients. The size of prediction errors associated with the simplifications assumed by the model are currently unknown.

The current paper represents a useful study to compute the actual capacity of the low region of LEO, based on the stability and sustainability of the space environment. Furthermore, the obtained outcomes will also be helpful to increase our awareness about a cautious use of LEO and to provide a tool for further analysis about new guidelines for post mission disposal and satellites end of life which would lead to an acceptable, sustainable and safe use of space.

Future work will involve the validation and verification of the model as well calibration of model coefficients. Moreover, several improvements for the source-sink model will be implemented, including additional species, new fluxes and species transitions, and a more accurate dynamic atmospheric density model. The aim of all these improvements is to obtain an increasingly realistic model to study LEO orbital capacity while retaining the simplicity and tractable advantages associated with the source-sink methodology.

\section*{Conflicts of Interest}
The authors declare no conflict of interest. 
\bibliographystyle{ieeetr}
\bibliography{manuscript_arxiv.bib}

\end{document}